\input ./style/arxiv-general.cfg
\documentclass[aap,MSNbibl,dvips]{arximspdf}
\makeatletter
   \@ifpackageloaded{graphicx}{}{\usepackage{graphicx}}
\makeatother
\usepackage{mathbh}

\doi{10.1214/14-AAP1058} 
\volume{25}
\issue{5}
\pubyear{2015}
\firstpage{2671}
\lastpage{2707}
\docsubty{FLA}

\makeatletter
\renewcommand{\P}{\mathbf{P}}
\newcommand{\rrvert}{\vert}
\newcommand{\llvert}{\vert}
\def\ito/{It\^o}
\renewcommand{\mid}{|}
\newcommand{\ulambda}{\underline{\lambda}}
\newcommand{\olambda}{\overline{\lambda}}
\newcommand{\oS}{\overline{S}}
\newcommand{\ub}{\underline{b}}
\newcommand{\ob}{\overline{b}}
\newcommand{\uS}{\underline{S}}
\newcommand{\og}{\overline{g}}
\newcommand{\ug}{\underline{g}}
\newcommand{\tS}{\tilde{S}}
\newcommand{\tW}{\tilde{W}}
\newcommand{\tmu}{\tilde{\mu}}
\newcommand{\tsigma}{\tilde{\sigma}}
\newcommand{\tphi}{\tilde{\varphi}}
\newcommand{\tc}{\tilde{c}}
\newcommand{\tZ}{\tilde{Z}}
\newcommand{\tV}{\tilde{V}}
\newcommand{\tdelta}{\tilde{\delta}}
\newcommand{\tbeta}{\tilde{\beta}}
\newcommand{\bphi}{\bar{\phi}}
\newcommand{\bc}{\bar{c}}
\newcommand{\bs}{\bar{s}}
\newcommand{\bV}{\bar{V}}
\newcommand{\I}[1]{\mathbh{1}_{(#1)}}
\def\real{{\mathbb R}}
\def\F{{\mathcal F}}
\def\cU{{\mathcal U}}
\def\cD{{\mathcal D}}
\def\cI{{\mathcal I}}
\newcommand{\tgo}{_{t\geq0}}
\def\cS{\mathcal{S}}
\let\phi\varphi
\newtheorem{theorem}{Theorem}[section]
\newtheorem{lemma}{Lemma}[section]
\newtheorem{proposition}{Proposition}[section]
\newproclaim{remark}{Remark}[section]
\newproclaim{definition}{Definition}[section]
\makeatother

\begin{document}
\begin{frontmatter}

\title{Shadow price in the power utility case\thanksref{T1}}
\runtitle{Shadow price in the power utility case}

\begin{aug}
\author[A]{\fnms{Attila}~\snm{Herczegh}\ead[label=e1]{prince@cs.elte.hu}}
\and
\author[A]{\fnms{Vilmos}~\snm{Prokaj}\corref{}\ead[label=e2]{prokaj@cs.elte.hu}}
\runauthor{A. Herczegh and V. Prokaj}
\affiliation{E\"otv\"os Lor\'and University}
\address[A]{Department of Probability Theory\\
\quad and Statistics\\
E\"otv\"os Lor\'and University\\
P\'azm\'any P\'eter s\'et\'any 1/C\\
1117 Budapest\\
Hungary\\
\printead{e1}\\
\phantom{E-mail: }\printead*{e2}}
\end{aug}
\thankstext{T1}{Supported by the European Union and the European
Social Fund
under the Grant agreement no.~T\'AMOP 4.2.1/B-09/KMR-2010-0003.}

\received{\smonth{12} \syear{2011}}
\revised{\smonth{8} \syear{2014}}

%
\begin{abstract}
We consider the problem of maximizing expected power utility from
consumption
over an infinite horizon in the Black--Scholes model with proportional
transaction costs, as studied in
Shreve and Soner
[\textit{Ann. Appl. Probab.}~\textbf{4} (1994) 609--692].

Similar to
Kallsen and Muhle-Karbe
[\textit{Ann. Appl. Probab.} \textbf{20} (2010) 1341--1358], we derive a shadow price,
that is, a frictionless price process with values in the bid-ask
spread which leads to the same optimal policy.
\end{abstract}

%
\begin{keyword}[class=AMS]
\kwd[Primary ]{91B28}
\kwd{91B16}
\kwd[; secondary ]{60H10}
\end{keyword}
\begin{keyword}
\kwd{Shadow price process}
\kwd{transaction costs}
\kwd{optimal consumption}
\kwd{power utility}
\end{keyword}
\end{frontmatter}

\section{Introduction}

It is a classical problem of mathematical finance to consider the
problem of
maximizing expected utility from consumption. This was initiated by
Merton \cite{Merton1969,MR0456373}, and thus is often referred to as the
Merton problem. He found
that for logarithmic or power utility it is optimal to keep a constant
fraction of wealth in stocks and to consume at a rate proportional to current
wealth.

This was extended to proportional transaction costs by
Magill and Constantinides \cite{MR0469196}. They stated
that it is optimal to restrain from
trading while the fraction of wealth invested in stocks is inside an interval
$[\theta_1,\theta_2]$. Their heuristic argument was made precise by
Davis and Norman \cite{MR1080472}, which was then generalized by Shreve and Soner \cite{MR1284980} who
managed to remove a couple of assumptions needed in \cite{MR1080472}.

These papers use methods from stochastic control. In recent years, it seems
there is more and more emphasis on solving portfolio optimization problems
with transaction costs by determining the shadow price of the problem;
see, for example,
Kallsen and Muhle-Karbe \cite{MR2676941},
Gerhold, Muhle-Karbe and Schachermayer \cite{2010arXiv10055105G},
Gerhold et al. \cite{2011arXiv11081167G}. This is a
process that establishes a link between portfolio
optimization with and without transaction costs as the optimal policy
of the
shadow price without frictions must coincide with that of the original problem.

The first article in this context is Kallsen and Muhle-Karbe \cite{MR2676941}. They use this
dual approach to come up with a free boundary
problem and solve that to derive the shadow price for logarithmic
utility. They also showed a connection with the original solution of
Davis and Norman \cite{MR1080472}. They point out how the optimal consumption derived
by Davis and
Norman can be used to determine the shadow value process and from that the
shadow price itself.

Our paper basically does the same for the power utility case.
In trying to apply the method of Kallsen and Muhle-Karbe \cite{MR2676941} to power utility,
one faces the problem that the optimal consumption plan of the
shadow market seems to be untractable. So we take a tour in optimal control
at a heuristic level. It provides an extra insight and finally a nontrivial
form of the consumption plan. Once we have this, we can carry out the analysis
similar to Kallsen and Muhle-Karbe \cite{MR2676941}.
Our main result is the following, all notions are defined in
Section~\ref{sec2}.

%
\begin{theorem}\label{thmmain}
Assume that the price $S$ is a geometric Brownian
motion
\[
dS_t=S_t(\mu \,dt+\sigma \,dW_t).
\]
The investor uses power utility $u(x)=x^\gamma/\gamma$, has
impatience rate $\delta>0$ and
faces proportional
transaction cost, that is she can
sell at $(1-\ulambda)S$ and buy at $(1+\olambda)S$, where
$\ulambda\in(0,1)$ and $\olambda>0$.

If
\[
\frac{\mu}{\sigma^2} \notin \{0,1-\gamma\}\quad\mbox{and}\quad \delta>
\frac{1}2\frac{\gamma}{1-\gamma}\frac{\mu^2}{\sigma^2}
\]
then there is a shadow price $\tS$ for the Merton problem for
sufficiently small transaction costs.

If, moreover,
\[
\mu<0\quad\mbox{or}\quad \delta>\gamma \biggl(\mu-\frac{\sigma^2}2(1-\gamma)
\biggr)
\]
then the shadow price exists for arbitrary $\ulambda\in(0,1)$ and
$\olambda>0$.
\end{theorem}

The rest of the paper is organized as follows. Section~\ref{sec2}
introduces the model
and summarizes some well-known result for the frictionless case.
Section~\ref{sec3}
contains heuristic arguments on how to come up with the
candidate for the shadow price.
Section~\ref{secshadowmarket}
analyzes the structure of the shadow market and ends with
the free boundary value problem, similar to that of obtained by
Kallsen and Muhle-Karbe~\cite{MR2676941}. The main new observation,
that makes it possible to carry out the analysis, is the form of the optimal
consumption. It uses the extra insight provided by the heuristics of the
optimal control approach. In Section~\ref{sec5},
we prove Theorem~\ref{thmmain}.
The elementary, however, painful and rather long,
analysis of the free boundary problem is
given in the \hyperref[app]{Appendix}. In Section~\ref{secasymp}, we compute the
asymptotic solution of the free boundary value problem and obtain the
asymptotic expansion for the no-trade region as well for the relative
consumption rate.

\section{Model and known results}\label{sec2}

\subsection{The model}

We study the problem of maximizing expected utility from consumption
over an
infinite horizon in the presence of proportional transaction costs as in
\cite{MR1080472,MR2589621,MR2676941,MR1284980}. We start with the
model description and define the basic notions.\vadjust{\goodbreak}

We consider a market with a bank account and a risky asset, a stock,
whose price evolution is given by
%
\begin{equation}
\label{eqdS} dS_t=S_t(\mu \,dt+\sigma
\,dW_t),
\end{equation}
with $S_0,\mu,\sigma>0$, where $W$ is a Brownian motion on the filtered
probability space $(\Omega,(\F_t)\tgo,\P)$. Whenever trading
occurs, the
investor faces higher ask (buying) and lower bid (selling) prices,
namely he
can buy at $\overline{S}_t=(1+\olambda)S_t$ and sell at
$\underline{S}_t=(1-\ulambda)S_t$ for some
$\olambda\in(0,\infty)$ and $\ulambda\in(0,1)$. Obviously, some
simplifications are possible. The value of $\sigma$ reflects the
time unit used [$\sigma^2$ is the variance of $\ln(S_1/S_0)$], one
can assume without restricting the generality that
$\sigma=1$; see also Remark~\ref{remsigma} below. Also we have three
prices, from which only the
bid and ask prices are used. Again without
restricting the generality, we may assume that $\olambda=0$ and
$S=\oS$.

%
\begin{definition}
A trading strategy $(\varphi^0_t,\varphi^1_t)_{t\geq0}$ is a
predictable process, where $\varphi^0_t$ and $\varphi^1_t$ denote
the number of units in the bond and the stock at time~$t$,
respectively. A
consumption\vspace*{1.5pt} rate process $(c_t)_{t\geq0}$ is a progressively
measurable process with nonnegative values.
We refer to $(\phi^0,\phi^1,c)$, that is, the trading strategy
$(\phi^0,\phi^1)$ and consumption rate $c$ together, as the
portfolio--consumption process.
\end{definition}

Recall that in the frictionless case a portfolio--consumption process is called
self-financing if
%
\begin{equation}
\label{eqself-financing} V_t=\phi^1S_t+
\phi^0=V_0+\int_0^t
\phi^1_s\,dS_s-\int_0^t
c_s\,ds.
\end{equation}

When the transaction cost is nonzero, transactions of infinite
variation lead to bankruptcy, so in this case we limit ourselves to
trading strategies of finite variation.
Then we can decompose
$\varphi^1_t=\varphi^{\uparrow}_t-\varphi^{\downarrow}_t$ as the
difference of the cumulative number of shares bought
$\varphi^{\uparrow}_t$ and sold $\varphi^{\downarrow}_t$ up to time
$t$.
We call a portfolio--consumption process self-financing, if
%
\begin{equation}
\label{eqsf} d\varphi^0_t= -\overline{S}_t\,d
\,\varphi^{\uparrow}_t+\underline{S}_t\,d\varphi
^{\downarrow}_t-c_t\,dt
\end{equation}
holds. Note, that when $\phi^1$ is of finite variation and
$\oS_t=\uS_t=S_t$ then we get back~(\ref{eqself-financing}), the
self-financing condition of the frictionless case.

Observe also, that in a self-financing portfolio--consumption process
$\phi^0$ is determined by $(\phi^1,c)$.

%
\begin{definition}
A self-financing portfolio--consumption process is admissible if its
liquidation value is nonnegative, that is,
\[
V_t^{\varphi}= \varphi^0_t+
\underline{S}_t\varphi^{+}_t-
\overline{S}_t\varphi ^{-}_t\geq0,\qquad
\mbox{a.s. for all }t\geq0.
\]
Given an initial endowment $(x_0,y_0)$, referring to the value of bonds and
stocks, respectively, the set of admissible strategies is denoted by
$\mathcal{A}(x_0,y_0)$.
We denote the value function by $v$, that is,
%
\begin{equation}
\label{eqv} v(x_0,y_0)=\sup_{(\varphi^1,c)\in\mathcal{A}(x_0,y_0)} {
\mathbf{E} \biggl(\int_0^{\infty}{e^{-\delta t}u(c_t)}
\,dt \biggr)}.
\end{equation}
Here, $\delta>0$ denotes a
fixed given impatience rate, $u$ a utility function.
\end{definition}

The goal of this paper is to identify the optimal
portfolio--consumption process for the market with transaction costs
as the optimal portfolio--consumption process of a suitably chosen
shadow market. This shadow market is frictionless and has the same
impatience parameter.

%
\begin{definition}
A shadow price (or rather the shadow problem) is a continuous
semi-martingale $\tS_t$, lying within the
bid-ask spread ($\uS_t\leq\tilde{S}_t\leq\oS_t$
a.s.), such that the optimal portfolio--consumption process for the
frictionless market with price $\tS$ is such that
it sells shares only when $\tS_t=\uS_t$ and buys them only when
$\tS_t=\oS_t$.
\end{definition}

Obviously, for any price process lying in the bid-ask spread, the
maximal expected
utility is at least as high as for the original market with price process
$S_t$, since the investor can trade at a smaller ask and a higher bid
price. Indeed, this is what makes the shadow price so special, the optimal
strategy with respect to it must only buy (resp., sell) when the shadow price
coincides with the original ask (resp., bid) price. We summarize this
observation in the following lemma.

\begin{lemma}
Assume that the shadow market with price $\tS$ and optimal
portfolio--consumption process $(\phi,c)$ exists.

If $(\phi,c)$ is admissible on the original market with
transaction cost, then it is optimal on this market as well.
\end{lemma}

The admissibility of $(\phi,c)$ may fail; see the discussion in
Section~\ref{sec42} below.

%
\begin{remark}\label{remsigma}Assume that we change the time unit so
that the volatility of $S$ becomes one. Then we have to re-scale $c$
and $\delta$ also; $c$ gives the consumption per unit time, and
$\delta$ is similar to a continuous interest rate.

More precisely,
assume that $S$ is a geometric Brownian motion
$dS_t=S_t(\mu \,dt+\sigma \,dW_t)$.
Consider the deterministic time-change $\eta(t)=\sigma^{-2} t$. Then
$\tW_t=\sigma W_{\eta(t)}$ is a Brownian motion and it generates
the filtration $(\F_{\eta(t)})_{t\geq0}$. The time-changed process
$S_{\eta(t)}$ is a geometric Brownian motion with parameters
$\tmu=\sigma^{-2}\mu$ and \mbox{$\tsigma=1$}. If $(\phi,c)$ is an admissible
self-financing portfolio--consumption process for the original
problem, then $\tphi_t=\tphi_{\eta(t)}$ and $\tc_t=\sigma^{-2}
c_{\eta(t)}$ constitute an admissible self-financing
portfolio--consumption\vspace*{1pt} process for the time-changed problem.
Finally, let $\tdelta=\sigma^{-2}\delta$. Then due to the form of
the power utility we have that
\[
\int_0^\infty e^{-\delta t} u(c_t)
\,dt= \sigma^{-2(1-\gamma)}\int_0^\infty
e^{-\tdelta t}u(\tc_t) \,dt.
\]
So it is enough to consider the case when $\sigma=1$; see also
\cite{MR942619}.
\end{remark}

\subsection{The problem without transaction costs}\label{secwocost}

The aim of this subsection is to formulate a
characterization of the optimal portfolio consumption process for the
power utility when the price of a stock $\tS$ follows an \ito/ process.
In the proof, we closely follow \cite{MR1121940}, Section~5.8. Even
though they only talk about finite time horizon,
the essence easily goes through to the infinite horizon case.

We start with a filtered probability space $(\Omega,(\F_t)\tgo,\P)$
and a
Brownian motion $W$, such that $\F$ is the filtration generated by $W$.
We assume that the discounted price process $\tS$ is an \ito/
process of the form
%
\begin{equation}
\label{eqtS} d\tS_t=\tS_t(\tmu_t\,dt+
\tsigma_t\,dW_t),
\end{equation}
where\vspace*{1pt}
$(\tmu_t)_{t\geq0}$ and $(\tsigma_t)_{t\geq0}$ are progressively measurable
and integrable, that is, $\int_0^t \llvert \tmu_s\rrvert +\tsigma
_s^2\,ds<\infty$
almost surely for all $t\geq0$.

We consider the Merton problem, that is, to find an admissible self-financing
port\-folio--con\-sump\-tion process that maximizes the expected
utility of the consumption
discounted by the impatience factor $\delta>0$ with a given initial endowment.
Beside the price process $\tS$ and the impatience factor $\delta$ we
fix a utility function $u$ which is assumed to be strictly increasing, concave
and two times \mbox{continuously} differentiable.

%
\begin{proposition}\label{proptZ}
Let $(\phi^0,\phi^1,c)$ be an admissible self-financing
portfo\-lio--consumption process.

If
%
\begin{equation}
\label{proptZit1} \bigl(e^{-\delta t} u'(c_t)
\bigr)_{t\geq0}\quad\mbox{and}\quad \bigl(e^{-\delta t}
u'(c_t)\tS_t \bigr)_{t\geq0}\qquad
\mbox{are local martingales}
\end{equation}
and
%
\begin{equation}
\label{eqV0} \mathbf{E} \biggl(\int_0^\infty
e^{-\delta t} u'(c_s) c_s\,ds \biggr)
= u'(c_0) V^{(\phi,c)}_0
\end{equation}
then $(\phi^0,\phi^1,c)$ is an optimal portfolio--consumption
process for the utility $u$.
\end{proposition}

Note, that since $\F$ is a Brownian filtration each $\F$-local martingale
has continuous sample paths.
\begin{pf*}{Proof of Proposition \ref{proptZ}}
Put
%
\begin{equation}
\label{eqZc} \tZ_t=e^{-\delta t}u'(c_t).
\end{equation}
By assumption, $\tZ$ and $\tZ\tS$ are nonnegative local martingales.

Let $(\bphi^0,\bphi^1,\bc)$ be an arbitrary admissible, self-financing
portfolio consumption process, starting from the same initial
endowment as $(\phi^0,\phi^1)$. Denote by
$\bV_t=V^{(\bphi,\bc)}_t=\bphi^1_t\tS_t+\bphi^0_t$
the value of this portfolio at time $t$. By admissibility and the
self-financing
assumption
\[
\bV\geq0,\qquad d\bV_t=\bphi^1_td
\tS_t-\bc_t \,dt.
\]
Application of the \ito/ formula yields that
%
\begin{equation}
\label{eqMdef} M_t:=\tZ_t\bV_t+\int
_0^t \tZ_s \bc_s\,ds=
\tZ_0 \bV_0+\int_0^t
\bphi^1_s\,d(\tZ\tS)_s +\int
_0^t \bphi ^0_s\,d
\tZ_s
\end{equation}
is also nonnegative local martingale. Let $(\tau_n)$ be a reducing
sequence of stopping times for the local martingale $M$. Since
$\tZ\bV\geq0$, we get
\[
\tZ_0\bV_0=M_{0}=\mathbf{E}(M_{\tau_n})
\geq\mathbf{E} \biggl(\int_0^{\tau_n}
\tZ_s \bc_s\,ds \biggr)\geq0.
\]
Letting $n\to\infty$, we get that for any admissible self-financing
portfolio
\begin{eqnarray*}
\mathbf{E} \biggl(\int_0^\infty\tZ_s
\bc_s\,ds \biggr)&\leq&\tZ_0\bV_0=\mathbf {E}
\biggl(\int_0^\infty\tZ_tc_t
\,dt \biggr).
\end{eqnarray*}

Since $u$ is concave, we have $u(\bc_t)-u(c_t)\leq
(\bc_t-c_t)u'(c_t)$ and
\[
\mathbf{E} \biggl(\int_0^\infty e^{-\delta t}
\bigl(u(\bc_t)-u(c_t) \bigr)\,dt \biggr)\leq \mathbf{E}
\biggl( \int_0^\infty(\bc_t-c_t)
\tZ_t \,dt \biggr)\leq0.
\]
\upqed
\end{pf*}
Under the assumptions of Proposition~\ref{proptZ}, the nonnegative
local martingale
\[
\tZ_tV^{(\phi,c)}_t+\int_0^t
\tZ_sc_s\,ds
\]
is
a closed martingale
%
\begin{equation}
\label{eqVfromQ} \tZ_tV^{(\phi,c)}_t=\mathbf{E}
\biggl( \int_t^\infty\tZ_sc_s
\Big| \F _t \biggr)\quad\mbox{and}\quad \lim_{t\to\infty}
\tZ_tV^{(\phi,c)}_t=0
\end{equation}
as it follows from the next statement.

\begin{proposition}\label{propM}
Let $M$ be a local martingale of the form
\[
M_t=\xi_t+\int_0^t
\psi_s\,ds,
\]
where $\xi$ and $\psi$ are nonnegative, adapted processes and
$\mathbf{E}(M_0)=\mathbf{E}(\int_0^\infty\psi_s\,ds)<\infty$. Then
\[
M_t=\mathbf{E} \biggl(\int_0^\infty
\psi_s\,ds \Big|\F_t \biggr)\quad\mbox{and}\quad
\xi_t=\mathbf{E} \biggl(\int_t^\infty
\psi_s\,ds \Big|\F_t \biggr).
\]
\end{proposition}

\begin{pf}
As $M$ is a nonnegative local martingale, it is a super-martingale, and
has a limit at infinity, say $M_\infty$. Then
\[
M_\infty\geq\int_0^\infty
\psi_s\,ds\quad\mbox{and}\quad \mathbf{E}(M_0)\geq
\mathbf{E}(M_\infty)\geq\mathbf{E} \biggl(\int_0^\infty
\psi_s\,ds \biggr)\geq\mathbf{E}(M_0).
\]
It follows that $M_\infty=\int_0^\infty\psi_s\,ds$ and
$\mathbf{E}(M_\infty\mid\F_t)=M_t$, from which the second part of
the claim
follows by subtracting $\int_0^t \psi_s\,ds$ from both sides.
\end{pf}
We add one more claim to this section which helps to check (\ref{eqV0}).

\begin{proposition}
\label{propconsumeall}
Assume that $(\phi^0,\phi^1,c)$ is an admissible, self-financing
portfolio--consumption process, such that
$(e^{-\delta t} u'(c_t))_{t\geq0}$ and
$(e^{-\delta t} u'(c_t)\tS_t)_{t\geq0}$ are local martingales.
Denote $\tV=V^{(\phi,c)}=\phi^1\tS+\phi^0$ the corresponding value process.

If
%
\begin{equation}
\label{eqsupZV} \mathbf{E} \Bigl(\sup_{t\geq0}e^{-\delta t}
u'(c_t)\tV_t \Bigr)<\infty\quad \mbox {and}
\quad e^{-\delta t} u'(c_t)\tV_t\to0
\qquad\mbox{a.s.}
\end{equation}
then (\ref{eqV0}) holds.
In particular, $(\phi^0,\phi^1,c)$ is an optimal
portfolio--consumption process.
\end{proposition}

\begin{pf}
Similar to the proof of Proposition~\ref{proptZ}, we use the
notation $\tZ_t=e^{-\delta t} u'(c_t)$.
Then as in (\ref{eqMdef})
the process
\[
\tZ_t\tV_t+\int_0^t
\tZ_s c_s\,ds
\]
is a local martingale.
Let $(\tau_n)_{n\geq1}$ be a reducing sequence of stopping times for
this local martingale. Then
\[
\tZ_0\tV_0=\mathbf{E} \biggl(\int_0^{\tau_n}
\tZ_s c_s\,ds \biggr)+\mathbf {E} \bigl((\tZ
\tV)_{\tau_n} \bigr).
\]
Letting $n\to\infty$ the second term goes to zero by the assumptions,
while the first one increases to $\mathbf{E}(\int_0^\infty\tZ
_sc_s\,ds)$ by
the monotone convergence theorem, hence the statement follows.
\end{pf}

The\vspace*{1pt} conditions (\ref{proptZit1}) and (\ref{eqV0}) formulated in
Proposition~\ref{proptZ}
not only are sufficient, but in some sense also necessary for
$(\phi^0,\phi^1,c)$ to be the optimal portfolio--consumption
process. Take the power utility as $u(x)=x^\gamma/\gamma$ and assume
that there is local martingale density $\tZ$, with $\tZ_0=1$ for
$\tS$. Now, define $c_t$ by inverting (\ref{eqZc}), that
is, $c_t=c_0 (e^{\delta t}\tZ_t)^{1/(\gamma-1)}$. If the left-hand
side of (\ref{eqV0}) is finite for some $c_0$, then there is a choice
of $c_0$ such that (\ref{eqV0}) holds\vspace*{1pt} and
one can find
a portfolio process $(\phi^0,\phi^1)$, such that
$(\phi^0,\phi^1,c)$ is the optimal portfolio--consumption process; see~\cite{MR1121940}, Section~5.8, for details.

\section{Candidate for the shadow price process}\label{sec3}
In this section, we argue at the heuristic level. Our aim is to introduce
the necessary notation and to motivate the relations among them. Then built
on these relations we construct the shadow price and the optimal
portfolio--consumption process in the next sections. It is based on the
solution of a free boundary problem, similar to the method of
Kallsen and Muhle-Karbe \cite{MR2676941}.

As before, we denote by $(\phi^0,\phi^1,c)$ an admissible self-financing
portfolio--consumption process. By the self-financing condition (\ref{eqsf}),
$\phi^0$ is determined by~$(\phi^1,c)$.

As usual, we define the value of a given position, at time $t$, as the
supremum of the achievable discounted utilities from future consumptions
given the past up to time $t$.
Due\vspace*{1pt} to the fact that the price process is Markovian,
this value depends only on the current state of the triple
$(\phi^0_t,\phi^1_t,S_t)$. Also, since the price is a geometric Brownian
motion it actually depends only on the wealth invested in stock and in bond,
that is on $(\phi^0_t,\phi^1_tS_t)$.
So we have the value as $v(\phi^0_t,\phi^1_tS_t)$,
where $v$ is defined by the formula (\ref{eqv}). We do not deal here with
such problems as the measurability of $v$ or its smoothness. We simply assume
in the following heuristic derivation that $v$ is smooth enough for all the
calculation we make.

Note that not all positions are possible, due to the admissibility
requirement. We denote by $\cS$ the solvency
cone, the admissible values for $(\phi^0_t,S_t\phi^1_t)$ for which the
liquidation value is still nonnegative:
\[
\cS= \bigl\{(x,y)\in\real^2\dvtx x+(1+\overline{\lambda})y\geq0
\mbox{ and } x+(1-\underline{\lambda})y\geq0 \bigr\}.
\]

To find the optimal
portfolio--consumption process the investor has to decide at each time
$t$ the
amount of trading and consumption in the next infinitesimal time interval.
Trading is only reasonable when its gain is nonnegative which
translates into
the requirement that there are no sellings\vspace*{1pt} when
$
v_y(\phi^0_t,S_t\phi^1_t)>(1-\ulambda)v_x(\phi^0_t,S_t\phi^1_t)
$
and no buying of stocks when
$v_y(\phi^0_t,\break S_t\phi^1_t)<(1+\olambda)v_x(\phi^0_t,S_t\phi^1_t)$. This
results in the existence of a nontrading region, if the investor behaves
rationally then no trading occurs when
\[
(1-\ulambda)v_x \bigl(\phi^0_t,S_t
\phi^1_t \bigr)< v_y \bigl(
\phi^0_t,S_t \phi^1_t
\bigr)< (1+\olambda)v_x \bigl( \phi^0_t,S_t
\phi^1_t \bigr).
\]
Similar analysis shows that
%
\begin{eqnarray}
\label{eqvxvy} && (1-\ulambda)v_x \bigl(\phi^0_t,S_t
\phi^1_t \bigr)\leq v_y \bigl(
\phi^0_t,S_t\phi^1_t
\bigr)\leq (1+\olambda)v_x \bigl(\phi^0_t,S_t
\phi^1_t \bigr)
\nonumber
\\[-8pt]
\\[-8pt]
\eqntext{\mbox{for all }t>0,}
\end{eqnarray}
as if this inequality is violated then the investor would immediately
re-balance his portfolio.

Concerning the consumption, if $c_t$ is the consumption rate at time
$t$, then
it gives $(u(c_t)-c_t v_x(\phi^0_t,S_t\phi^1_t))\,dt $ gain\vspace*{1pt} in the next
infinitesimal
interval. It is maximal if $c_t=I(v_x(\phi^0_t,S_t\phi^1_t))$, where
$I=(u')^{-1}$.
So the for the optimal policy $(\bphi^0,\bphi^1,\bc)$
we must have
%
\begin{equation}
\label{eqvxu} v_x \bigl(\bphi^0_t,S_t
\bphi^1_t \bigr)=u'(\bc_t).
\end{equation}

For an admissible, self-financing portfolio--consumption process
$(\phi,c)$ we have that
%
\begin{equation}
e^{-\delta t} v \bigl(\phi^0_t,S_t
\phi^1_t \bigr)+\int_0^t
e^{-\delta s} u(c_s)\,ds
\end{equation}
is a super-martingale, as on each time-interval $[t,t+\Delta t]$ the
average decrease of the first term exceeds the average gain of the
second given the past $\F_t$, by the very definition of $v$.
For the optimal strategy $(\bphi^0,\bphi^1,\bc)$,
the expectation has to be constant, yielding the characterization that
$(\bphi^0,\bphi^1,\bc)$ is optimal exactly when
%
\begin{equation}
\label{eqDPP} e^{-\delta t} v \bigl(\bphi^0_t,S_t
\bphi^1_t \bigr)+\int_0^t
e^{-\delta s} u(\bc_s)\,ds
\end{equation}
is a martingale.

For $\gamma>0$, all terms appearing in this martingale
are positive, while for $\gamma<0$ they are negative.
Next, we treat the case
$\gamma\in(0,1)$. For $\gamma<0$, our conclusion also holds, but one has
to multiply by $-1$ the whole expression and then repeat the same argument.

So we fix $\gamma\in(0,1)$ and denote by $(\bphi,\bc)$ the optimal
strategy. Then for the martingale given in (\ref{eqDPP}), Proposition
\ref{propM} yields that
%
\begin{equation}
\label{eqintu|Ft} e^{-\delta t}v \bigl(\phi^0_t,S_t
\phi^1_t \bigr)= \mathbf{E} \biggl(\int
_t^\infty e^{-\delta s} u(\bc_s)\,ds
\Big|\F_t \biggr).
\end{equation}

Assume now that there is a shadow market with a shadow price $\tS$
such that
$(\bphi^0,\bphi^1,\bc)$ is the optimal portfolio--consumption process
for $\tS$
without transaction cost satisfying the conditions of Proposition
\ref{proptZ}.
Then by (\ref{eqvxu})
\[
\tZ_t=e^{-\delta t}u'(\bc_t)
=e^{-\delta t} v_x \bigl(\bphi^0_t,
\bphi^1_tS_t \bigr),
\]
and due to the structure of the power utility
$u(x)=x^\gamma/\gamma$
we also have that $u'(x)x=\gamma u(x)$. That is, the right-hand side of
(\ref{eqintu|Ft}) can be written using (\ref{eqVfromQ}) as
%
\begin{equation}
\label{eqvtVtZ} e^{-\delta t}v \bigl(\phi^0_t,S_t
\phi^1_t \bigr)= \frac{1}{\gamma}\mathbf{E} \biggl(\int
_t^\infty\tZ_s \bc_s\,ds \Big|
\F_t \biggr)= \frac{1}{\gamma}\tV_t\tZ_t.
\end{equation}
As $\tV_t=\bphi^1_t\tS_t+\bphi^0_t$, the shadow price $\tS$ must satisfy
%
\begin{equation}
\label{eqtZtVfromv} \tZ_t\tS_t=\frac{\tZ_t\tV_t-\tZ_t\bphi^0_t}{\bphi^1_t}
=e^{-\delta t}S_t\cdot\frac{\gamma
v(x,y)-v_x(x,y)x}{y} \bigg|_{x=\bphi^0_t,y=\bphi^1_tS_t}.
\end{equation}
Since $(\phi^1_t,c_t)\in\mathcal{A}(x,y)$ exactly when $
(\alpha\phi^1_t,\alpha c_t)\in\mathcal{A}(\alpha x,\alpha y)$ for any
$\alpha>0$, the value function $v$ is homothetic $v(\alpha x,\alpha
y)=\alpha^{\gamma}v(x,y)$. That is, we can write
%
\begin{equation}
\label{eqhomthx+y} v(x,y)=(x+y)^{\gamma}h \biggl(\frac{x}{x+y} \biggr),
\end{equation}
with $h(z)=v(z,1-z)$.
The homotheticity of the value function formalizes the intuition that only
the proportion of the wealth held in shares is relevant.

From (\ref{eqhomthx+y}), we obtain by easy calculation that in the
domain of
$v$
%
\begin{equation}
\label{eqvvxvy} \gamma v(x,y) = xv_x(x,y)+yv_y(x,y).
\end{equation}
Writing it back to (\ref{eqtZtVfromv}), we obtain the main formula of this
heuristic derivation. If shadow price exists then it can be expressed
with the
optimal portfolio--consumption process $(\bphi^0,\bphi^1,\bc)$ and
the value
function $v$ in the form
%
\begin{equation}
\label{eqtSmrs} \tS_t=\frac{v_y(\bphi^0_t,\bphi^1_tS_t)}{v_x(\bphi^0_t,\bphi
^1_tS_t)} S_t,
\end{equation}
that is the shadow price is the marginal rate of substitution. Note
that $\tS$
lies in the bid-ask spread due to (\ref{eqvxvy}).

The same observation was made in the
case of power utility and for the problem of maximizing terminal wealth
in a
finite time horizon in \cite{Mark2000209}. More recently, this
connection was also found in \cite{2011arXiv11081167G},
for the power utility, but for the optimal growth rate problem without
consumption.

One could show at this point, using the results of Shreve and Soner \cite{MR1284980}
(see also
the recent monograph \cite{MR2589621} for the exposition of their
result), that
$\tS$ is indeed a shadow price. They
showed that the value function for this problem is smooth enough and
satisfies the so-called smooth pasting conditions. Then the martingale
property of $\tZ$ and $\tZ\tS$ as defined above follows easily, and
$\tS$ is
the shadow price for the problem. However, their work is based on the
viscosity solution of the Hamilton--Jacobi--Bellman equation, while
our method is rather elementary.

Our original motivation stem from the paper of Kallsen and Muhle-Karbe \cite{MR2676941}, where
the logarithmic utility was treated and similar treatment of power utility
was posed. The rest of the paper is
devoted to this; we identify the
functional identities implied by heuristics, and from this we obtain a
free boundary value problem, very similar to that of \cite{MR2676941},
we analyse this ODE, and from its solution we finally construct the
shadow market.

\section{Structure of the shadow market}\label{secshadowmarket}
Combination of (\ref{eqtSmrs}) and the homotheticity
(\ref{eqhomthx+y}) of $v$ easily yields
that the ratio $\tS/S$ should depend only on\vspace*{1pt} the proportion of the
wealth invested in bond. The same is true for the ratio
$\bc_t/(\phi^0_t+\phi^1_tS_t)$ the relative consumption rate.
This suggests that there is a fundamental process $\tbeta$ behind the scene
and all relevant information can be obtained from it. In the case of
logarithmic utility,\vspace*{1pt} the same idea was applied by
Kallsen and Muhle-Karbe \cite{MR2676941} and
their analysis is based on the process
$\ln\llvert  {\bphi^1_t\tS_t}/{\bphi^0_t}\rrvert$. Gerhold et al. \cite{2011arXiv11081167G} uses
the normalized version $\ln\llvert {\bphi^1_t S_t}/{\bphi
^0_t}\rrvert $ in
the power
utility case, without consumption, that is, they use the price $S$
instead of
the shadow price $\tS$.
The two approaches are equivalent. To cover the case when the no-trade region
is not disjoint from the axes, we use a third variant.

In our problem,
the real difficulty is
the complicated form of the optimal consumption. In \cite{MR2676941},
the fact that the optimal consumption in the shadow
market is a fixed proportion of the wealth counted with the shadow price
simplified the analysis greatly while in \cite{2011arXiv11081167G}
there is
no consumption.

In the rest of this section, we describe the structure of our shadow
market. We
return to the notation $(\phi^0,\phi^1,c)$ for a portfolio consumption
process, $\tS$ is the price of a share in this market, while $S$, the
price on
the market with transaction cost, is a geometric
Brownian motion as in (\ref{eqdS}). We choose the time unit such that
the volatility of $S$ is one.
Based on the above heuristics, we seek the shadow price candidate $\tS
$ and the
relative consumption rate as smooth functions of $\tbeta$ which is
assumed to
be a reflected diffusion in an interval $I$. $I$ can be a semi-closed
$(\ub,\ob]$ or $[\ub,+\ob)$ or a bounded closed interval
$[\ub,\ob]$, that is,
%
\begin{equation}
\label{eqtbeta} d\tbeta_t=\mu_{\tbeta}(\tbeta_t)
\,dt+ \sigma_{\tbeta}(\tbeta_t)\,dW_t+dL_t,
\end{equation}
where the bounded variation process $L$ keeps the diffusion $\tbeta$ in
$I$ and satisfies $\int_0^t
\I{\tbeta_s\in\partial I}\,d L_s=L_t$ for all $t\geq0$ almost
surely. Here\vspace*{1.5pt} and in what follows, $\partial I$ denotes the boundary
points of~$I$ contained in~$I$. When an end point of the interval~$I$ is not
included in $I$, then it means that the process does not reach this end
point. It will turn out that in our parameterization
$I$ is either closed or of the form $[\ub,0)$.

Note that, under quite general conditions on the coefficients in
(\ref{eqtbeta}) the solution of the SDE (\ref{eqtbeta}) exists,
unique in law and $\tbeta$ is a Markov process; see,
for example,~\cite{MR0145598}.
In the\vspace*{1pt} following general recipe of the shadow market, the concrete
meaning of
$\tbeta$ is not relevant.

We define $\tS$ in the form
%
\begin{equation}
\label{eqtSfromg} \tS_t=S_t\exp\bigl\{g(\tbeta_t)\bigr\},
\end{equation}
where $g\dvtx I\to\real$ is a $C^2$ function such that
$g'\mid_{\partial I}=0$. This boundary condition guarantees
that $\tS$ is an \ito/ process, as the effect of the singular part $d
L_t$ is
annulled. Since $\tS$ is a positive \ito/ process, we
write the evolution of $\tS$ as
%
\begin{equation}
\label{eqdtS} d\tS_t=\tS_t \bigl(\tmu(
\tbeta_t)\,dt+\tsigma(\tbeta_t)\,dW_t \bigr),
\end{equation}
where
%
\begin{equation}
\label{eqtmutsigma} \tsigma=1+g'\sigma_{\tbeta},\qquad \tmu-
\tfrac{1}2\tsigma^2=\mu-\tfrac{1}2+g'
\mu_{\tbeta}+\tfrac{1}2 g''
\sigma^2_{\tbeta}
\end{equation}
by \ito/'s formula.

We will use $\alpha\dvtx I\to\real$ for the function which expresses the
proportion of the wealth held in stocks
in terms of $\tbeta_t$, that is,
we think of $\alpha(\tbeta_t)$ as
$\phi^1_t\tS_t/(\phi^0_t+\phi^1_t\tS_t)=\phi^1_t\tS_t/\tV_t$.

Finally, we will use the
notation $\rho$ for the function which expresses the relative
consumption rate
from $\tbeta$, that is, we think of $\rho(\tbeta_t)$ as $c_t/\tV_t$.

So we have to chose the interval $I$ the functions
$\mu_{\tbeta},\sigma_{\tbeta},g,\rho,\alpha\dvtx I\to\real$, and the initial
value $\tbeta_0$ for the process $\tbeta$. Then with the
solution of the SDE
(\ref{eqtbeta})
we can define $\tV_t$ as the stochastic exponential of
$\alpha(\tbeta_t)\,d\tS_t/\tS_t-\rho(\tbeta_t)\,dt$. Here, the
first term is $\phi^1_t/\tV_t\,d\tS_t$ while the second is $-c_t/\tV
_t\,dt$, so the
definition of $\tV$ follows the identity (\ref{eqself-financing}). Formally,
we define $\tV$ as
%
\begin{equation}
\label{eqtVfromtbeta} \tV_t=\tV_0\exp\biggl\{\int_0^t
\frac{\alpha(\tbeta_t)}{\tS_t}\,d \tS_t- \int_0^t
\frac{1}2 \frac{\alpha^2(\tbeta_t)}{\tS^2_t}\,d \langle\tS \rangle_t-\int
_0^t\rho(\tbeta_t)\,dt\biggr\},
\end{equation}
where $\tV_0=\phi^0_0+\tS_0\phi^1_0$. $\phi^0_0$ and $\phi^1_0$
are the number
of bonds and stocks at time zero.

From $\tV$, $\tS$ and $\tbeta$ we can define the
portfolio--consumption process
$(\phi^0,\phi^1,c)$ as
%
\begin{equation}
\label{eq01c} \phi^1_t=\alpha(\tbeta_t)
\frac{\tV_t}{\tS_t},\qquad \phi^0_t= \bigl(1-\alpha(
\tbeta_t) \bigr)\tV_t,\qquad c_t=\rho(
\tbeta_t)\tV_t.
\end{equation}

Before going on, note that any choice of the interval
$I$ and of the smooth functions
$\mu_{\tbeta},\sigma_{\tbeta},g,\rho,\alpha$, satisfying some
regularity conditions,
such that $\rho\geq0$
leads to a system of processes through the equations
(\ref{eqtbeta}), (\ref{eqtSfromg}), (\ref{eqtVfromtbeta}) and
(\ref{eq01c}), and these processes satisfy by definition
%
\begin{equation}
\alpha(\tbeta_t)=\frac{\phi^1_t\tS_t}{\tV_t}, \qquad \tV_t =
\phi^1_t\tS_t+\phi^0_t,
\qquad d\tV_t=\phi^1_t \,d
\tS_t-c_t \,dt,
\end{equation}
that is the self-financing condition holds. The process $\tV_t> 0$ for all
$t\geq0$, and $c_t\geq0$, hence $(\phi^0,\phi^1,c)$ is an admissible
self-financing
portfolio--consumption process for the market with price $\tS$. The price
process $\tS$ is an \ito/ process provided that the boundary condition
$g'\mid_{\partial I} =0$ holds; moreover, $\ln(\tS/S)$ evolves in the
range of $g$.
Note also, that although $\phi^0$, $\phi^1$, $c$
are all diffusions, $(\phi^0,\phi^1)$ are not necessarily of bounded
variation.

The choice of $\alpha$ determines the meaning of $\tbeta$.
We will use the identity function~$\alpha$, that is
$\tbeta_t=\phi^1_t\tS_t/(\phi^0_t+\phi^1_t\tS_t)$.\vspace*{6pt}

\textit{Notation.} To shorten the
notation, we write $\tmu_t$ and $\rho_t$ for $\tmu(\tbeta_t)$ and
$\rho(\tbeta_t)$, respectively, and similarly for
other functions of the process $\tbeta_t$.

\subsection{Trading when \texorpdfstring{$\tilde{\beta}$}{tildebeta} is extremal}

In order to find the shadow price,
we have to chose $\mu_{\tbeta},\sigma_{\tbeta},g,\rho$. Not all
choices will result in a shadow market. Here,\vspace*{-1pt} we make a new
assumption, namely that $g$ is monotone, then $\tS_t/S_t$ is
extremal if and only if $\tbeta_t\in\partial I$. The next proposition shows
that the requirement that trading is allowed, that is, $\phi^1$ can change,
only when $\tbeta$ hits the
boundary of $I$ implies some nontrivial relations among
$\mu_{\tbeta},\sigma_{\tbeta},g,\rho$ and $\alpha$.


\begin{proposition}\label{propedge}
Let $\tbeta,\tV,\tS,\phi^0,\phi^1,c$ be a solution of the system of
equations (\ref{eqtbeta}), (\ref{eqtSfromg}), (\ref{eqtVfromtbeta}) and
(\ref{eq01c}).
The number of shares $\phi^1$ changes only when the process $\tbeta$ is
at the boundary of $I$, if and only if
%
\begin{equation}
\label{eqtmutsigmatbeta} \qquad\sigma_{\tbeta}=\alpha(1-\alpha)\tsigma\quad
\mbox{and}\quad \mu_{\tbeta}=\alpha(1-\alpha) \bigl(\tmu-\alpha
\tsigma^2 \bigr)+ \alpha\rho \qquad\mbox{holds on $I$}.
\end{equation}

In this case,
\begin{equation}
\label{eqphi^1fromtbeta} \phi^1_t=\phi^1_0
e^{\int_0^t 1/\alpha_s \,dL_s}\quad\mbox{and}\quad d\phi^0_t=
\tV_t(-dL_s-\rho_t\,dt).
\end{equation}
\end{proposition}

\begin{pf} The idea\vspace*{1pt} is that
$\phi^1$ changes only when $\tbeta_t$ is extremal if and only if
the evolution of
$\ln\llvert \phi^1_t\rrvert=\ln\llvert  \alpha_t\rrvert -(\ln
\tS_t-\ln\tV_t)$ is
driven by the singular part~$dL_t$ of $d\tbeta_t$. Note that $\ln\llvert \alpha\rrvert '=1/\alpha
$ and
$\ln\llvert \alpha\rrvert ''=-1/\alpha^2$ so
%
\begin{equation}
\label{eqdlnalpha} d\ln\llvert \alpha_t\rrvert= \frac{1}{\alpha_t}\,d
\tbeta_t-\frac{1}{2\alpha_t^2}\,d\langle\tbeta \rangle_t.
\end{equation}
The dynamics of $(\ln\tS_t-\ln\tV_t)$
%
\begin{eqnarray}
\label{eqStVt} && \ln\tS_t-\ln \tV_t
\nonumber
\\
&&\qquad = \biggl(\tmu_t-\frac{\tsigma_t^2}2 \biggr)\,dt+
\tsigma_t\,dW_t- \biggl(\alpha_t
\tmu_t-\frac{1}2 \alpha_t^2
\tsigma^2_t \biggr)\,dt - \alpha _t
\tsigma_t\,dW_t +\rho_t\,dt\hspace*{-15pt}
\\
&&\qquad = \biggl((1-\alpha_t)\tmu_t-\frac{1}2
\bigl(1- \alpha^2_t \bigr)\tsigma^2_t+
\rho_t \biggr)\,dt+ (1-\alpha_t)\tsigma_t
\,dW_t.
\nonumber
\end{eqnarray}
Now $\ln\llvert \phi^1\rrvert $ is driven by $dL_t$ exactly when
the drift
and diffusion terms in
(\ref{eqdlnalpha}) and~(\ref{eqStVt}) are equal.
By the assumed regularity of $\tbeta$, it is equivalent to the functional
identity (\ref{eqtmutsigmatbeta}).

We also obtained that (\ref{eqtmutsigmatbeta}) implies
$d\ln\llvert \phi^1_t\rrvert=1/\alpha_t\,dL_t$ which proves the
first part of
(\ref{eqphi^1fromtbeta}). The second part follows from the self-financing
condition
$d\phi^0_t=-\tS_t\,d\phi^1_t-c_t\,dt=
\tV_t(-\alpha_t\,d\ln\llvert \phi^1_t\rrvert-\rho_t\,dt)$.
\end{pf}

If trading happens only when $\tbeta\in\partial I$, that is
(\ref{eqtmutsigmatbeta}) holds, then we can replace the identities
in (\ref{eqtmutsigma}) with more convenient ODE-s for $\tsigma$ and
$g$. These equations will be used later.

%
\begin{proposition}\label{propfromtsigma}
Consider the next two equations:
%
\begin{eqnarray}
\label{eqode} \frac{1}2 \alpha(1-\alpha)^2
\tsigma'&=& (1-\alpha) \biggl(\frac{1}{\tsigma} \biggl(\tmu-
\frac{1}2\tsigma^2 \biggr)- \biggl(\mu -\frac{1}2
\biggr) \biggr)-\frac{\tsigma-1}{\tsigma}\rho,
\\
\label{eqgfromtsigma} \alpha(1-\alpha)g'\tsigma&=&\tsigma-1.
\end{eqnarray}
Assume that (\ref{eqtmutsigmatbeta}) holds, that is, trading happens
only when the $\tbeta_t\in\partial I$. Then
(\ref{eqode}) and (\ref{eqgfromtsigma}) together are equivalent to
(\ref{eqtmutsigma}).
\end{proposition}

\begin{pf}
The first part of (\ref{eqtmutsigma}),
(\ref{eqtmutsigmatbeta}) and (\ref{eqgfromtsigma}) can be
written as
\[
\tsigma-1=g'\sigma_{\tbeta},\qquad \sigma_{\tbeta}=
\alpha(1-\alpha)\tsigma \quad\mbox{and}\quad \tsigma-1 =\alpha(1-\alpha)
g'\tsigma,
\]
respectively.
Obviously, the first two of these equations imply the third and the
last two
imply the first.
This shows that when (\ref{eqtmutsigmatbeta}), that is, the
identity in the middle holds, then the first and last identities
are equivalent.

Note that (\ref{eqgfromtsigma}) claims that $1$ the constant
volatility of $S$ factorizes as
\[
1= \bigl(1-\alpha(1-\alpha) g' \bigr)\tsigma.
\]
Hence, $\tsigma$ is nonzero on $I$, which is also implicitly
contained in
(\ref{eqode}).

Now assume that (\ref{eqtmutsigma}), (\ref{eqtmutsigmatbeta}) hold. Then
we have (\ref{eqgfromtsigma}) and two expressions for~$\mu_{\tbeta
}$. We show
that the comparison of these two formulas yields the ODE for
$\tsigma$ in (\ref{eqode}).

By (\ref{eqgfromtsigma}), we get
\[
\tsigma=\frac{1}{1-\alpha(1-\alpha)g'}\quad\mbox{and}\quad \tsigma' =-
\tsigma^2 \bigl((2\alpha-1) g'-\alpha(1-
\alpha)g'' \bigr).
\]
Recall, that $\alpha$ denotes the identity function on $I$.
Now $\alpha(1-\alpha)g'\tsigma=\tsigma-1$ and
$\alpha^2(1-\alpha)^2\tsigma^2=\sigma_{\tbeta}^2$ so
we obtain that
\[
\alpha(1-\alpha)\tsigma'= -(2\alpha-1)\tsigma(\tsigma-1)+
\sigma_{\tbeta}^2 g''
\]
that is
%
\begin{equation}
\label{eqg} \frac{1}2 g''
\sigma_{\tbeta}^2 = \frac{\tsigma-1}{\tsigma} \biggl(\alpha-
\frac{1}2 \biggr) \tsigma^2+ \frac{1}2\alpha(1-\alpha)
\tsigma'.
\end{equation}
Now the second half of (\ref{eqtmutsigmatbeta}) yields using (\ref
{eqgfromtsigma})
\begin{eqnarray}
\label{eqg1} %
(1-\alpha)g' \mu_{\tbeta} &=&
\alpha(1-\alpha)g' \bigl((1-\alpha) \bigl[ \tmu-\alpha
\tsigma^2 \bigr]+\rho \bigr)
\nonumber
\\[-8pt]
\\[-8pt]
\nonumber
&=& \frac{\tsigma-1}{\tsigma} \bigl((1-\alpha) \bigl(\tmu-\alpha
\tsigma^2 \bigr)+\rho \bigr).
\end{eqnarray}
So we get
\begin{eqnarray}\label{eq1-alphagg}
&& (1-\alpha) \biggl(g'\mu_{\tbeta}+
\frac{1}2 g''\sigma_{\tbeta}^2
\biggr)
\nonumber\\[-8pt]\\[-8pt]\nonumber
&&\qquad = \frac{\tsigma-1}{\tsigma} \biggl((1-\alpha) \biggl[ \tmu-\frac{1}2
\tsigma ^2 \biggr]+\rho \biggr)+ \frac{1}2\alpha(1-
\alpha)^2 \tsigma'.
\end{eqnarray}
By substituting (\ref{eq1-alphagg}) into the identity obtained
by multiplying the second
part of~(\ref{eqtmutsigma}) with $(1-\alpha)$,
we get (\ref{eqode}).

Conversely, (\ref{eqtmutsigmatbeta}) and (\ref{eqgfromtsigma})
implies (\ref{eq1-alphagg}). Then (\ref{eqode}) is just $(1-\alpha)$
times the second half of (\ref{eqtmutsigma}). Since $(1-\alpha)\neq
0$ on
$I\setminus\{1\}$ we obtain that the second half of (\ref
{eqtmutsigma}) holds on the whole $I$ by continuity.
\end{pf}

\subsection{\texorpdfstring{$\tilde{S}$}{tildeS} as the marginal rate of substitution}\label{sec43}
We have seen in Section~\ref{sec3} the shadow price must be the
marginal rate of substitution when one
uses power utility. It also imposes
some nontrivial relation among our functions.
To be precise, we take the analog of the value function of Section~\ref{sec3} based on the formula (\ref{eqvtVtZ}).
Assume that at time $t$ the state process $\tbeta_t=b$ and
$\tV_t=\phi^0_t+\phi^1_t\tS_t=V$. By\vspace*{1pt} formula~(\ref{eqvtVtZ}), the
value of our future consumption, that is, $v(\phi^0_t,\phi^1_tS_t)$
can be obtained as $u'(c_t)\tV_t$. In other words, it can be expressed
from $b$ and $V$. This expression, apart from the constant multiplier,
is given as a function of $b\in I$ and $V\geq0$ by the formula
$\rho(b)^{\gamma-1}V^\gamma$,
that is, we take $\tilde{v}\dvtx I\times\real_+\to\real$ the function
expressing the
value of the position (without constant factors) in terms of
$\tbeta_t$ and $\tV_t$, as
\[
\tilde{v}(b,V)=\rho(b)^{\gamma-1}V^\gamma
\]
and $q\dvtx I\times\real_+\to\real^2$ the function which expresses
$(\phi^0_t,S_t\phi^1_t)$ in terms of $\tbeta_t$ and~$\tV_t$, that is,
\[
q(b,V)= \bigl((1-b)V,e^{-g(b)}bV \bigr).
\]
Then $\tS_t/S_t=e^{g(\tbeta_t)}$ must be the ratio of the partial
derivatives of $\tilde{v}\circ q^{-1}$ evaluated at $q(\tbeta_t,\tV_t)$.
We obtain by easy calculation that
$\tS$ is the marginal rate of substitution if and only
if
\[
-(\gamma-1)\rho'\alpha+\gamma\rho \bigl(1-\alpha g'
\bigr)= (\gamma-1)\rho'(1-\alpha)+\gamma\rho.
\]
We summarize this in the next proposition.

%
\begin{proposition}\label{propmrs}
Let $\tbeta,\tV,\tS,\phi^0,\phi^1,c$ be a solution of the system of
equations (\ref{eqtbeta}), (\ref{eqtSfromg}), (\ref{eqtVfromtbeta}) and
(\ref{eq01c}) and the condition (\ref{eqtmutsigmatbeta}) in
Proposition~\ref{propedge} hold.

Then the price $\tS$ is the marginal rate of substitution with respect to
$\tilde{v}$, that is,
%
\begin{equation}
\label{eqtStv} \tS_t=\frac{\tilde{v}_y(\phi^0_t,\phi^1_tS_t)}{\tilde{v}_x(\phi
^0_t,\phi^1_tS_t)}S_t
\end{equation}
if and only if
%
\begin{equation}
\label{eqlnrho} (\gamma-1) (\ln\rho)'=-\gamma\alpha
g' \qquad\mbox{on $I$}.
\end{equation}
\end{proposition}

\subsection{$c$ as the optimal consumption plan}

We still work in the framework introduced in Section~\ref{secshadowmarket}. That is,\vspace*{1pt}
we assume that $\tsigma,\tmu,\rho\dvtx I\to\real$ are smooth functions,
(\ref{eqtmutsigmatbeta}), (\ref{eqlnrho})
hold and the
processes $\tbeta$, $\tS$, $\tV$, $\phi^0$, $\phi^1$ and $c$ are
determined by
equations~(\ref{eqtbeta}), (\ref{eqtSfromg}), (\ref{eqtVfromtbeta}) and
(\ref{eq01c}).

Now we want to find conditions in terms of $g,\rho$ and $\alpha$ ensuring
that $(\phi^0,\phi^1,c)$ is the optimal portfolio--consumption
process. As
before, we translate (\ref{proptZit1}) of Proposition~\ref{proptZ}
into a
functional identity. One can write the equations that are dictated by
Proposition~\ref{proptZ}, however, it seems to be untractable without the
insight provided by the heuristics in Section~\ref{sec3} and
formulated in
(\ref{eqlnrho}).

%
\begin{proposition}\label{proprho} Let $\tbeta,\tV,\tS,\phi^0,\phi
^1,c$ be a
solution of the system of
equations (\ref{eqtbeta}), (\ref{eqtSfromg}), (\ref{eqtVfromtbeta})
and (\ref{eq01c}) and assume (\ref{eqtmutsigmatbeta}) and
(\ref{eqlnrho}).\vspace*{1pt}

Then $e^{-\delta t}c_t^{\gamma-1}$ and $e^{-\delta t}c_t^{\gamma
-1}\tS_t$ are local
martingales if and only if
%
\begin{eqnarray}
\label{eqlnrho40} \frac\tmu{\tsigma} &=& \alpha(\tsigma-\gamma),
\\
\label{eqrho} \rho&=&\frac{\delta}{1-\gamma}+ \frac{\alpha\gamma}{\gamma-1} \biggl(\mu-
\frac{1}2+\frac{1}2 \bigl[ (1-\alpha)\tsigma+\alpha\gamma \bigr]
\biggr).
\end{eqnarray}
\end{proposition}

\begin{pf} First note that, since $g'\mid_{\partial I}=0$ and
$(\ln\rho)'=-\frac{\gamma}{\gamma-1}\alpha g'$ by (\ref{eqlnrho}),
$\rho(\tbeta_t)$ is an \ito/ process.
Write $Z_t=e^{-\delta t} (c_t/c_0)^{\gamma-1}$ as
\[
Z_t=\exp\biggl\{\int_0^t a(
\tbeta_t) \,dW_t +\int_0^t
b(\tbeta_t) \,dt\biggr\}. %
\]
Then $Z_t$ is a
local martingale if and only if $b=- a^2/2$ and in this case
$dZ_t=Z_t a(\tbeta_t)\,dW_t$. Assuming this, the other process $Z_t\tS
_t$ is
a local martingale if and only if $a=-\tmu/\tsigma$.
So we have to express $a,b$ and check these conditions, taking the
identities (\ref{eqtmutsigmatbeta}) and (\ref{eqlnrho})
for granted.
Since $c_t=\exp\{\ln\rho(\tbeta_t)+\ln\tV_t\}$, we have the following
identities
\begin{eqnarray*}
a&=&(\gamma-1) \bigl((\ln\rho)'\sigma_{\tbeta}+\alpha\tsigma
\bigr),
\\
b&=&-\delta+(\gamma-1) \bigl((\ln\rho)'\mu_{\tbeta}+
\tfrac{1}2(\ln\rho )''\sigma^2_{\tbeta}+
\alpha\tmu-\tfrac{1}2\alpha^2\tsigma^2-\rho
\bigr).
\end{eqnarray*}

Using the relations $\sigma_{\tbeta}=\alpha(1-\alpha)\tsigma$
from Proposition~\ref{propedge} and $(\gamma-1)(\ln\rho)'=-\gamma
\alpha g'$
from Proposition~\ref{propmrs}, we get
\[
(\gamma-1) (\ln\rho)'\sigma_{\tbeta} =-\gamma(1-\alpha)
\alpha ^2g'\tsigma,
\]
and using also $(1-\alpha(1-\alpha)g')\tsigma=1$ from Proposition~\ref{propedge}
\begin{eqnarray*}
a &=&(\gamma-1) (\ln\rho)'\sigma_{\tbeta}+(\gamma-1)\alpha
\tsigma =-\gamma(1-\alpha)\alpha^2 g'\tsigma+(\gamma-1)
\alpha\tsigma
\\
&=&\gamma\alpha \bigl(1-\alpha(1-\alpha)g' \bigr)\tsigma-\alpha
\tsigma= \alpha(\gamma-\tsigma).
\end{eqnarray*}
This shows that $a=-\tmu/\tsigma$ exactly
when
(\ref{eqlnrho40}) holds.

To express $b+\frac{1}2 a^2$ we use again that
$(\gamma-1)(\ln\rho)'=-\gamma\alpha g'$, which gives
$(\gamma-1)(\ln\rho)''=-\gamma(\alpha g''+ g')$ as $\alpha'=1$. Hence,
\begin{eqnarray*}
(\gamma-1) (\ln\rho)''\sigma_{\tbeta}^2&=&
-\gamma \bigl(\alpha g''+g' \bigr)
\sigma^2_{\tbeta},
\\
(\gamma-1) (\ln\rho)'\mu_{\tbeta}&=& -\gamma\alpha
g'\mu_{\tbeta}.
\end{eqnarray*}
By (\ref{eqtmutsigma}),
\begin{eqnarray*}
g'\mu_{\tbeta}+\frac{1}2g''
\sigma_{\tbeta}^2&=& \biggl(\tmu-\frac{\tsigma^2}2 \biggr)-
\biggl(\mu-\frac{1}2 \biggr),
\\
g'\sigma_{\tbeta}&=&\tsigma-1.
\end{eqnarray*}
Using also that $\sigma_{\tbeta}=\alpha(1-\alpha)\tsigma$,
we get
\begin{eqnarray*}
&& (\gamma-1) \biggl((\ln\rho)'\mu_{\tbeta}+
\frac{1}2( \ln\rho)''\sigma
^2_{\tbeta} \biggr)
\\
&&\qquad = \gamma\alpha \biggl[ \biggl(\mu-\frac{1}2 \biggr) - \biggl(
\tmu- \frac{\tsigma^2}2 \biggr)- \frac
{(1-\alpha)(\tsigma-1)\tsigma}2 \biggr]
\\
&&\qquad = \gamma\alpha \biggl[ \biggl(\mu-\frac{1}2 \biggr) - \biggl(
\tmu- \alpha \frac{\tsigma^2}2 \biggr)+ (1-\alpha)\frac{\tsigma}2 \biggr].
\end{eqnarray*}
Then we have that
%
\begin{eqnarray} \label{eqb+a22}
&&b+\frac{a^2}2+\delta+(\gamma-1)\rho\nonumber
\\
&&\qquad = \gamma\alpha \biggl[ \biggl(\mu-\frac{1}2 \biggr) - \biggl(
\tmu- \alpha \frac{\tsigma^2}2 \biggr)+ (1-\alpha)\frac{\tsigma}2 \biggr]
\nonumber\\[-8pt]\\[-8pt]\nonumber
&&\quad\qquad{} + (\gamma-1)\alpha \biggl(\tmu-\alpha\frac{\tsigma^2}2 \biggr)+ \frac{\tmu^2}{2\tsigma^2}
\\
&&\qquad = \gamma\alpha \biggl(\mu-\frac{1}2+ (1-\alpha)
\frac{\tsigma}2 \biggr)-\alpha \biggl(\tmu-\alpha\frac{\tsigma^2}2 \biggr)+
\frac{\tmu
^2}{2\tsigma^2}.\nonumber
\end{eqnarray}
The last two terms can be expressed using (\ref{eqlnrho40}) as
\[
\frac{\tmu^2}{2\tsigma^2}-\alpha \biggl(\tmu-\alpha\frac{\tsigma^2}2 \biggr)=
\frac{1}2\alpha^2(\tsigma-\gamma)^2-\alpha \biggl(
\alpha\tsigma(\tsigma -\gamma)-\alpha\frac{\tsigma^2}2 \biggr)=
\frac{1}2 \alpha^2\gamma^2.
\]
Whence $b+a^2/2=0$ holds exactly when (\ref{eqrho}).
\end{pf}

\section{Synthesis}\label{sec5}
We have collected all the necessary relations among the unknown
functions. In the \hyperref[app]{Appendix}, we prove the existence of the pair
$(I,\tsigma)$ such that $\tsigma$ nowhere vanishing continuous
function on an interval $I$ satisfying the following ODE with boundary
condition:
%
\begin{eqnarray}\label{eqODEagain}
\frac{1}2\alpha(1-\alpha)^2
\tsigma'&=& (1-\alpha) \biggl(\frac{1}\tsigma \biggl(\tmu-
\frac{\tsigma^2}2 \biggr)- \biggl(\mu-\frac{1}2 \biggr) \biggr)-
\frac{\tsigma-1}\tsigma\rho,
\nonumber\\[-8pt]\\[-10pt]\nonumber
\tsigma\mid_{\partial I}&=&1,
\end{eqnarray}
where
%
\begin{eqnarray}\label{eqtmurhokappa}
\tmu &=&\tsigma\alpha(\tsigma-\gamma),\qquad \rho=
\frac{\delta}{1-\gamma}+ \frac{\alpha\gamma}{\gamma
-1}\kappa,
\nonumber\\[-8pt]\\[-8pt]\nonumber
\kappa&=&\mu-\frac{1}2+
\frac{1}2 \bigl((1-\alpha)\tsigma+\gamma\alpha \bigr),
\end{eqnarray}
$\alpha$ is the identity on $I$, and $\kappa$ is an auxiliary
notation, used also below in the proof of Proposition~\ref{proprho>0}.

To be more precise, in the \hyperref[app]{Appendix} we prove the existence of
$\tsigma$ for sufficiently small transaction costs under the condition
%
\begin{equation}
\label{eqdelta>gamma} \delta\geq\frac{1}2 \frac{\gamma}{1-\gamma}\mu^2,
\qquad\mu \notin\{0,1-\gamma\}.
\end{equation}
We also give more restrictive conditions in Theorems~\ref{thmb},
\ref{thmc},~\ref{thmd} for the existence of
the solution for any transaction cost.

There are two cases $\mu>0$ then $I\subset(0,\infty)$ is a closed
interval or $\mu<0$ then $I\subset(-\infty,0)$ and it may happen that
$I$ is not closed, but in this case $I$ has the form $[\ub,0)$ with
some $\ub<0$.

In what follows, we assume that for the given $\ulambda\in(0,1)$ and
$\olambda>1$ there is
$(I,\tsigma)$
such that
\[
\int_I \biggl\llvert \frac{\tsigma-1}{\tsigma\alpha(1-\alpha)} \biggr\rrvert =
\ln \frac{1+\olambda}{1-\ulambda},
\]
and the $\frac{\tsigma-1}{1-\alpha}$ is continuous and not vanishing on
$I$, in particular its sign is constant.

Given $\tsigma$ we define all other functions on $I$ in the natural
way: $\tmu,\rho$ by (\ref{eqtmurhokappa}) and $g$ as the integral
of
%
\begin{equation}
\label{eqgdef} g'=\frac{\tsigma-1}{\tsigma\alpha(1-\alpha)}
\end{equation}
such that its range is subset of $[\ln(1-\ulambda),\ln(1+\ulambda
)]$. Then $g$
is continuously differentiable and its second derivative exists and
continuous except may be at $1$. Then $\mu_{\tbeta}$ $\sigma_{\tbeta
}$ are
defined by the formula (\ref{eqtmutsigmatbeta}). By Proposition~\ref{propfromtsigma}, the relations among
$\tmu,\tsigma,\mu_{\tbeta},\sigma_{\tbeta},g$ given in (\ref
{eqtmutsigma}) hold.

\subsection{Shadow market}\label{sec51}
The functions $\mu_{\tbeta},\sigma_{\tbeta}$ are Lipschitz
continuous, so
when $I$ is a closed interval, equation
(\ref{eqtbeta}) defining $\tbeta$ has a unique strong solution for
any initial value by a
classical result of Skorohod \cite{MR0145598}.

When $I=[\ub,0)$, then first we
consider the equation for $\xi_t=\ln\llvert  \tbeta_t\rrvert $,
that is, we define
$\xi$ from the equation
%
\begin{equation}
\label{eqxi} d\xi_t=\sigma_\xi(\xi_t)
\,dW_t+ \mu_\xi(\xi_t)\,dt+L^{\xi}_t,
\end{equation}
where
\begin{eqnarray*}
\sigma_\xi(y)&=&\frac{\sigma_{\tbeta}}{\alpha} \bigl(-e^y \bigr),
\\
\mu_\xi(y)&=&\frac{\mu_{\tbeta}}{\alpha} \bigl(e^y \bigr)-
\frac{1}2\sigma_\xi^2(y),
\end{eqnarray*}
and $L^\xi_t$ is a process of bounded variation forcing the process
$\xi$ to be in $(-\infty,\ln\llvert  \ub\rrvert]$. The
coefficients for
$\xi$
are bounded and locally Lipschitz continuous, hence the solution of
(\ref{eqxi}) is unique and strong. Then $\tbeta_t=-\exp\{\xi_t\}$ is
the unique solution of (\ref{eqtbeta}).

Still for the case $I=[\ob,0)$ we remark that $\tbeta$ visits the
point $\ub$ infinitely often, that is, for each $t>0$ there are visits
after $t$ almost surely. Then by the strong Markov property of
$\tbeta$, it follows easily that $\int_0^\infty\tbeta_t^2 \,dt=\infty$
almost surely.

We\vspace*{1pt} define the shadow price process from the state process $\tbeta$
which is a
reflected diffusion on $I$ as it is described in Section~\ref{secshadowmarket}.
When $\tbeta,\tV$ is the state of the shadow market then
on the original market the value of the bank
account is $(1-\tbeta)\tV$ and the value of shares is $\tbeta
e^{-g(\tbeta)}\tV$. That is, the
no-trade region introduced in Section~\ref{sec3} is the interior of
the cone
\[
\bigl\{ \bigl((1-b)V,be^{-g(b)}V \bigr)\dvtx b\in I, V>0 \bigr\}.
\]
At time $t=0$, we are given
$\phi^0_{0^-}$ the number of bonds and $\phi^1_{0^-}$ the number of
shares and $S_0$.
It may
happen that our initial position is not in the closure of the no-trade region.
In this\vspace*{1.5pt} case, we have to re-balance our position to achieve
this and set $(\tbeta_0,\tV_0)$ to be the corresponding point in
$I\times[0,\infty)$.

When $\mu<0$ and $\frac{1+\olambda}{1-\ulambda}$ is large enough, then
it may happen that $(\tbeta_0,\tV_0)$ obtained in this way is such
that $\tbeta_0=0$. It means that we have no shares at time zero and we
do not buy as the price is a strict super-martingale. Then the price
$\tS$ has no role as there is no trading involved in the optimal
strategy. In what follows, we deal with the case when $\tbeta_0\in I$.

So the construction described in Section~\ref{sec3} yields
$\tS,\tV,\phi^0,\phi^1,c$. Then $(\phi^0,\phi^1,c)$ is an admissible
self-financing portfolio--consumption process.\vspace*{1pt}
Note that admissibility here means admissibility with respect to the
price $\tS$.

\subsection{Regularity of the state process \texorpdfstring{$\tilde{\beta}$}{tildebeta}}
When $I$ is disjoint from the set $\{0,1\}$, then the
coefficients of
equation (\ref{eqtbeta}) are bounded and
$\sigma_{\tbeta}=\alpha(1-\alpha)\tsigma$ is also bounded from
below. The regularity in this case is obvious, that is, $\tbeta$ visits
all points of $I$ whatever is the initial value.
Then $\tbeta$ hits both endpoints of $I$ and
selling and buying of shares occurs infinitely often.

Regularity is also rather straightforward, when $I=[\ub,0)$ as in this
case the previous properties hold for $\xi=\ln\llvert  \tbeta
\rrvert$. In this
case, $\tbeta$ hits $\ub$ infinitely often, but never hits $0$. In
terms of trading, it means that we start with negative number of
shares, and when the prices go too low we buy them, realizing the
profit of our short position.

There\vspace*{1pt} is, however, the case when
$I=[\ub,\ob]$ contains $1$. This can happen when $\mu>(1-\gamma)$ and
the transaction costs are high, more precisely
$\ln\frac{1+\olambda}{1-\ulambda}$ is large enough. Then
$\sigma_{\tbeta}(1)=0$ and $\mu_{\tbeta}(1)>0$. It implies that
$\tbeta$
will reach $1$ in finite time if it started from below, and
immediately enters to the region $(1,\ob]$. This position means that
we take debt on the bank account to finance our consumption but keep
the number of shares. By comparison of solutions with different
starting values\vspace*{1pt} (for details see
\cite{revuz-yor}, Chapter IX, Theorem~3.7), one can easily show
that when $\tbeta$ entered into $(1,\ob]$ it stays
there forever, meaning that we have negative value on the bank account
and when the share price goes high we realize the profit by selling
some shares.

\subsection{Optimility of \texorpdfstring{$(\phi,c)$}{(phi,c)} on the frictionless market with price \texorpdfstring{$\tilde{S}$}{tildeS}}
Next, we want to check that $(\phi,c)$, defined above in Section~\ref{sec51},
is the
optimal portfolio--consumption process for the price $\tS$ on a market
without transaction costs. For that, we use Proposition~\ref{proptZ},
that is, we need to show (\ref{proptZit1}) and (\ref{eqV0}).

With the notation $\tZ=e^{-\delta t}
c_t^{\gamma-1}$ condition (\ref{proptZit1}) requires that both $\tZ$ and
$\tZ\tS$ are local martingales.
Since (\ref{eqlnrho40}) and (\ref{eqrho}) were used to define $\tmu$
and $\rho$ in (\ref{eqtmurhokappa}) all, but~(\ref{eqlnrho}) of
the conditions of Proposition~\ref{proprho} holds obviously.
Equation~(\ref{eqlnrho}) is the
relation
\[
(\gamma-1) (\ln\rho)'=-\gamma\alpha g'.
\]
As\vspace*{1pt} $g$ is defined through (\ref{eqgdef}), the next
proposition claims that (\ref{eqlnrho}) also holds and, therefore, by
Proposition~\ref{proprho} $\tZ$ and $\tZ\tS$ are local martingales.

\begin{proposition}\label{proprho>0}
Let $\tsigma,\tmu,\rho,\kappa\dvtx I\to\real$ be continuous functions
such that $\tsigma$ is nowhere vanishing and satisfies the ODE
(\ref{eqODEagain}) and
(\ref{eqtmurhokappa}) holds.
Then
\[
(1-\alpha) (\gamma-1)\rho'=-\gamma\frac{\tsigma-1}{\tsigma}\rho.
\]
In particular, by Proposition~\ref{proprho} $\tZ$ and $\tZ\tS$ are
local martingales.
\end{proposition}

\begin{pf}
$\tmu,\rho,\kappa$ are differentiable by (\ref{eqtmurhokappa})
and we have
\[
\delta+(\gamma-1)\rho=\gamma\alpha\kappa.
\]
Therefore,
%
\begin{equation}
\label{eqrhokappa} (1-\alpha) (\gamma-1)\rho' =\gamma(1-\alpha) \bigl(
\kappa+\alpha \kappa' \bigr).
\end{equation}
Again by (\ref{eqtmurhokappa}),
\[
(1-\alpha) \bigl(\kappa+\alpha\kappa' \bigr)= \frac{1}2
\alpha(1-\alpha)^2\tsigma' -(1-\alpha) \biggl(
\frac{1}{\tsigma} \biggl[ \tmu-\frac{\tsigma^2}{2} \biggr]- \biggl[ \mu -
\frac{1}2 \biggr] \biggr).
\]
Using (\ref{eqODEagain}), we get
\[
(1-\alpha) \bigl(\kappa+\alpha\kappa' \bigr)=-\frac{\tsigma-1}{\tsigma
}
\rho.
\]
So the right-hand side of (\ref{eqrhokappa}) simplifies to
$-\gamma\frac{\tsigma-1}{\tsigma}\rho$ and the claim follows.
\end{pf}

The next proposition shows that (\ref{eqV0}) is also fulfilled and completes
the proof of the optimality of
$(\phi,c)$. When $I$ is not contiguous to 0, then $\alpha^2$ is bounded
from below, while for $I=[\ub,0)$ we have already remarked that
$\int_0^\infty\alpha_t^2\,dt=\infty$ almost surely. So in each case
$\int_0^\infty\alpha_t^2\,dt=\infty$.
Recall the notation
$f_t=f(\tbeta_t)$ for the process obtained from the state process
$\tbeta$.

\begin{proposition}\label{proplim=0}
If $\int_0^\infty\alpha^2_t=\infty$, then
\[
\mathbf{E} \Bigl(\sup_{t\geq0}\tZ_t\tV_t
\Bigr)<\infty\quad\mbox{and}\quad \tZ_t\tV_t\to0 \qquad
\mbox{a.s.}
\]
\end{proposition}
\begin{pf}
In the proof of
Proposition~\ref{proprho}, we obtained the dynamics of
$\tZ_t=e^{-\delta t} c_t^{\gamma-1}$ is given by $d\tZ_t=
(-\tmu_t/\tsigma_t) \tZ_t\,dW_t$ provided that our set of functions
satisfies (\ref{eqlnrho}), (\ref{eqlnrho40}) and (\ref{eqrho}). We have seen that all these
identities hold in our construction. Using also (\ref{eqtVfromtbeta})
we have
\[
d \bigl(\ln(\tZ_t\tV_t) \bigr)= \alpha_t(
\tmu_t\,dt+\tsigma_t\,dW_t)-\frac{1}2
\alpha^2_t\tsigma ^2\,dt-\rho_t
\,dt- \frac{\tmu_t}{\tsigma_t} \,dW_t -\frac{\tmu^2_t}{2\tsigma^2_t}\,dt.
\]
Here, $\frac{\tmu}{\tsigma}=\alpha(\tsigma-\gamma)$ by (\ref
{eqlnrho}) and the expression simplifies to
\[
d \bigl(\ln(Z_t\tV_t) \bigr)=\gamma\alpha_t
\,dW_t- \bigl(\rho_t+\tfrac{1}2(\gamma
\alpha_t)^2 \bigr)\,dt.
\]
Since the function $\rho>0$ is continuous on $I$, $\alpha$ is the
identity on $I$
and $I$ is bounded we also have that $2\rho/\alpha^2$ is bounded
from below, denote by $\eta>0$ a lower bound. Then $\ln(Z_t\tV
_t)\leq M_t-\frac{1}2(1+\eta)\langle M\rangle_t$ with a continuous local
martingale $M$ whose dynamics is
$dM_t=\gamma\alpha_t \,dW_t$ and $M_0=\tZ_0\tV_0$.
Using the well-known estimate for the tail
probability of the supremum of a Brownian motion with negative drift, we
get first $\P(\sup_t \ln(\tZ_t\tV_t) > r)\leq e^{-2(1+\eta)r}$
and then $\mathbf{E}(\sup_t \tZ_t\tV_t)<\infty$.
As $\langle M\rangle_\infty=\infty$ we also have
$M_t-\frac{1}2(1+\eta)\langle M\rangle_t\to-\infty$ almost
surely which gives $Z_t\tV_t=e^{-\delta t}c_t^{\gamma-1}\tV_t\to0$.
\end{pf}

We have proved that $(\phi,c)$ is the optimal portfolio--consumption
process on the frictionless market with price $\tS$.

\subsection{Admissibility of \texorpdfstring{$(\phi^0,\phi^1,c)$}{(phi0,phi1,c)} under the price $S$}\label{sec42}
We have seen that $(\phi^0,\phi^1,c)$ is an admissible self-financing
portfolio--consumption process for the price $\tS$.
With $\ug=\inf_I g$ and $\og=\sup_I g$, we have $1-\ulambda=e^{\ug
}$ and
$1+\olambda=e^{\og}$. Then the liquidation value of the portfolio in the
market with proportional transaction costs is the minimum of the next
two expressions
\begin{eqnarray*}
\phi^0_t+\phi^1_t
\uS_t&=& \bigl((1-\alpha_t)+\alpha_te^{-g_t}e^{\ug
}\bigr)\tV_t,
\\
\phi^0_t+\phi^1_t
\oS_t&=& \bigl((1-\alpha_t)+\alpha_te^{-g_t}e^{\og
}\bigr)\tV_t.
\end{eqnarray*}
As $\tV_0>0$ and, therefore, $\tV_t>0$ for all $t\geq0$, admissibility
holds exactly when
%
\begin{eqnarray}
\label{eqad} e^{\og-g}\alpha+(1-\alpha)&\geq&0,\qquad e^{\ug-g}
\alpha+(1-\alpha)\geq0\qquad\mbox{on }I.
\end{eqnarray}

The admissibility of $(\phi^0,\phi^1,c)$ with respect to $S$ is
obvious if
$0<\mu<1-\gamma$ as in this case $I\subset(0,1)$. In other words,
the wealth held in shares and on the bank account are both positive,
therefore, so is the liquidation value.

The other cases are not so trivial. When $\mu>(1-\gamma)$, then
$I\subset(0,\infty)$ so $\alpha>0$ on $I=[\ub,\ob]$ and the admissibility
condition simplifies to
\[
\frac{1-\ulambda}{1+\olambda}\ob+1-\ob\geq0\quad\iff\quad \ob\leq\frac{1+\olambda}{\olambda+\ulambda}.
\]
Here, $\ob$ is obtained from the solution of the free boundary problem.
If $\olambda,\ulambda\to0$, then the corresponding $\ob$ converges to
$\mu/(1-\gamma)$. That $(\phi,c)$ is admissible when the transaction
costs are small enough.
The explanation is that when the
transaction cost increases the no trading region is increases and at
the same
time the solvency cone shrink to the positive orthant. So for a large
transaction cost it happens that even the Merton line lies outside the
solvency cone.

For $\mu<0$, our conclusion is similar. The admissibility condition
simplifies to
\[
\frac{1+\olambda}{1-\ulambda}\ub+1-\ub\geq0\quad\iff\quad \ub\geq-\frac{1+\olambda}{\olambda+\ulambda}.
\]

\section{Asymptotics}\label{secasymp}
Similar to \cite{MR2995515}, we can derive the asymptotic expansion of
the boundaries and we compare these to \cite{MR2048827}.
In this section, we compute the asymptotic solution of the free
boundary value problem.

In the \hyperref[app]{Appendix}, we prove that under the
condition
%
\begin{equation}
\label{eqdelta>=} \delta\geq\frac{1}2\frac{\gamma}{1-\gamma} \mu^2
\end{equation}
the free boundary value problem has a solution $(I,f)$ for
sufficiently small transaction
costs; the solution is defined on $I=[x,s(x)]$ where
$x<x_0=\mu/(1-\gamma)$ and $s(x)=\inf\{y>x\dvtx f(y)=1\}$. More
precisely, $f\dvtx I\to(0,\infty)$ solves
%
\begin{equation}
\label{eqfODErecall} \tfrac{1}2 f'=a_0f+(1-f)
\bigl((a_1+a_2)f+a_3 f^2 \bigr),
\qquad f\mid_{\partial I}=1.
\end{equation}
Then $\tsigma=1/f$ solves the ODE (\ref{eqODEagain}) on $I$. In the
asymptotic analysis, the only important properties of the function
coefficients $a_0,a_1,a_2,a_3$ are that they are analytic around $x_0$
and $a_0(x_0)=0$, while $a'_0(x_0)\neq0$. The concrete form of this
functions are given below in (\ref{eqadef}).

Let us introduce the function $h_z(y)=f_{x_0-z}(x_0+yz)$. For small
$z$ and $x_0=\mu/(1-\gamma)\notin\{0,1\}$, the function
$h_z$ will be defined on $[-1,2]$ and solves the integral equation
\[
h_z(y)=1+2z \int_{-1}^y F
\bigl(x_0+zu,h_z(u) \bigr)\,du,
\]
where $F(\cdot ,f)$ is the right-hand side of the ODE in (\ref
{eqfODErecall}). Then the Taylor expansion of the two variable function
$(z,y)\mapsto h_z(y)$ takes the form
\[
h_z(y)=\sum_{k\geq0} z^k
p_k(y).
\]
If we denote by $[ z^k]$ the operator which takes the coefficient of
$z^k$ in the Taylor expansion of an analytic function, we get the
following recursion for $(p_k)_{k\geq0}$:
\[
p_0(y)=1,\qquad p_{k}(y)= 2\int_{-1}^y
\bigl[ z^{k-1} \bigr]F \bigl(x_0+zu,h^{[k-1]}_z(u)
\bigr)\,du, \qquad k\geq1,
\]
where
\[
h^{[n]}_z(y)=\sum_{0\leq k\leq n}
z^k p_k(y).
\]

The first few terms of the approximation of $h_z$ are easily computed
and all other terms are computable, for example,
\begin{eqnarray*}
p_1(y)&=&0,
\\
p_2(y)&=&a'_0(x_0)
\bigl(y^2-1 \bigr)=\frac{(1-\gamma)^3}{\mu(1-\gamma-\mu
)} \bigl(1-y^2 \bigr),
\\
p_3(y)&=&\frac{1}3a_0''(x_0)
\bigl(y^3+1 \bigr)+\frac{2}3a_0'(x_0)
(a_1+a_2+a_3) (x_0)
\bigl(y^3-y \bigr).
\end{eqnarray*}
The impatience parameter $\delta$ appears in $p_3$ only, through the
value of $a_3(x_0)$, similar to
the remark in \cite{MR2995515}.

Once we have the expansion of $h_z$, we get that
$\bs(z)=\inf\{y>-1\dvtx h_z(y)=1\}$ also admits an expansion around
zero and its coefficients can be calculated recursively. More precisely,
we take the alternative definition of $\bs(z)$ as
\[
\bs(z)=\inf \biggl\{y>-1\dvtx\sum_{k\geq0}
z^k p_{k+2}(y)=0 \biggr\}.
\]
Nothing has changed for $z>0$, but it has no jump at $z=0$ and
gives $\bs(0)=1$.
The first few terms of the expansion of $\bs$ are
\begin{eqnarray*}
\bs(0)&=&1,\qquad\bs'(0)=\frac{p_3(1)}{p_2'(1)}=\frac
{a_0''(x_0)}{3a'_0(x_0)}.
\end{eqnarray*}

Then
\begin{eqnarray*}
\cI(x_0-z)&=&\int_{x_0-z}^{s(x_0-z)} \biggl
\llvert \frac{f_{x_0-z}(y)-1}{y(1-y)} \biggr\rrvert \,dy\Big|_{y=x_0+zu}
\\
&=& \int_{-1}^{\bs(z)} \biggl\llvert
\frac{h_z(u)-1}{(x_0+zu)(1-x_0-zu)} \biggr\rrvert z\,du
\\
&=& z^3\frac{4}3 \biggl\llvert \frac{a'_0(x_0)}{x_0(1-x_0)} \biggr
\rrvert +O \bigl(z^4 \bigr).
\end{eqnarray*}
Higher-order expansion is also possible, since the integrand does not
change sign for small $z$. However, we content ourself with the first
nonzero term of the expansion.
In the formula above, $a_0'(x_0)=(1-\gamma) (x_0(1-x_0))^{-1}$, so
\[
\frac{a_0'(x_0)}{x_0(1-x_0)}=\frac{1-\gamma}{x_0^2(1-x_0)^{2}}>0.
\]
Recall that here $x_0$ is
the Merton proportion $x_0=\mu/(1-\gamma)$.

To get the asymptotics for the size of the no-trade region, we measure
the transaction cost with a single number
$\lambda=\frac{\olambda+\ulambda}{1+\olambda}$. Then
$\uS=(1-\lambda)\oS$ and
%
\begin{eqnarray}\label{eqzlambda}
&& \cI \bigl(x_0-z(\lambda) \bigr)=\ln\frac{1}{1-\lambda}
\nonumber\\[-8pt]\\[-8pt]\nonumber
&&\quad\iff\quad z(\lambda)= \biggl(\frac{3}4 {\frac{x_0(1-x_0)}{a'_0(x_0)}}
\biggr)^{1/3} \lambda ^{1/3}+O \bigl(\lambda^{2/3}
\bigr).\nonumber
\end{eqnarray}
Since $s(x_0-z)=x_0+z\bs(z)=x_0+z(1+O(z))$, we have that for small
$\lambda$ the solution of the ODE (\ref{eqfODErecall})
is defined on $I=[\ub(\lambda),\ob(\lambda)]$ with
\begin{eqnarray*}
\ub&=&x_0- \biggl(\frac{3}4\frac{x_0^2(1-x_0)^2}{(1-\gamma)}
\biggr)^{1/3} \lambda^{1/3}+O \bigl(\lambda^{2/3}
\bigr),
\\
\ob&=&x_0+ \biggl(\frac{3}4\frac{x_0^2(1-x_0)^2}{(1-\gamma)}
\biggr)^{1/3} \lambda^{1/3}+O \bigl(\lambda^{2/3}
\bigr).
\end{eqnarray*}
From this, the result of Jane{\v{c}}ek and Shreve follows easily. They
considered the case when strict inequality holds in (\ref{eqdelta>=}).
For a given $\lambda$, consider the function
\[
\theta_\lambda(x)=\frac{xe^{-g_\lambda(x)}}{(1-x)+xe^{-g_\lambda(x)}},
\]
where $g_\lambda$ is the function belonging to the transaction cost
$\lambda$.
$\theta_\lambda$ gives the
proportion of wealth held in shares when it counted with the price $S$
given that the proportion counted with $\tS$ is $x$ and the
transaction cost is $\lambda$. Then $\theta_\lambda$ is differentiable
and $\lim_{\lambda\to0^+}\theta'_\lambda(x_0)=1$. It can be obtained
by direct calculation, but also clear from the meaning of
$\theta_\lambda$. So for small $\lambda$ we have that
\begin{eqnarray*}
\theta_\lambda(\ub)&=&x_0- \biggl(\frac{3}4
\frac{x_0^2(1-x_0)^2}{(1-\gamma)} \biggr)^{1/3} \lambda^{1/3}+O \bigl(
\lambda^{2/3} \bigr),
\\
\theta_\lambda(\ob)&=&x_0+ \biggl(\frac{3}4
\frac{x_0^2(1-x_0)^2}{(1-\gamma
)} \biggr)^{1/3} \lambda^{1/3}+O \bigl(
\lambda^{2/3} \bigr).
\end{eqnarray*}
The careful reader may realize that the constant is half of the one in
Jane{\v{c}}ek and Shreve \cite{MR2048827}, Theorem~2. The reason is that our $\lambda$ is
twice of the $\lambda$ used in that paper.

We also compute the expansion of the consumption rate. Again we only
compute the first nonzero correction term. Similarly as above, we
start with a solution of the ODE $(f_{x_0-z},I)$.
Then in the corresponding shadow market, the relative consumption
rate is given by the function
%
\begin{equation}
\label{eqrhoz} \rho_z(y)=\frac{\delta}{1-\gamma}+\frac{\gamma y}{\gamma-1}
\biggl(\mu-\frac{1}2+\frac{1}2 \biggl[ (1-y)\frac{1}{f_{x_0-z}(y)}+y
\gamma \biggr] \biggr).
\end{equation}
As we are interested in the shape of $\rho_z$ for small $z$ we re-scale
it to
\[
r_z(u)=\rho_z(x_0+uz), \qquad u\geq-1.
\]
When $z=0$ (\ref{eqrhoz}) simplifies the well-known value
of optimal relative consumption rate for the frictionless case
\[
\rho_0(x_0)=\frac{\delta}{1-\gamma}-\frac{\gamma}{2(1-\gamma
)^2}
\mu^2= \frac{\delta}{1-\gamma}-\frac{\gamma}{2}x_0^2,
\]
and $r_0$ is the constant function taking this value.
Then one gets
\begin{eqnarray*}
\frac{\gamma-1}{\gamma} \bigl(r_z(u)-r_0(u) \bigr)&=&
\frac{(x_0+uz)(1-(x_0+uz))}2 \biggl(\frac{1}{h_z(u)}-1 \biggr)-\frac
{1-\gamma}2(uz)^2
\\
&=& \frac{x_0(1-x_0)}2 \bigl(h_z(u)-1 \bigr)-\frac{1-\gamma}2(uz)^2+O
\bigl(z^3 \bigr)
\\
&=& \frac{x_0(1-x_0)}2 \bigl(z^2p_2(u) \bigr)-
\frac{1-\gamma}2(uz)^2+O \bigl(z^3 \bigr)
\\
&=& -\frac{1-\gamma}2 z^2+ O \bigl(z^3 \bigr).
\end{eqnarray*}
This formula says that the first correction term due to the friction
is a constant change of the consumption rate. Plugging in
(\ref{eqzlambda}), we get the next approximation of the optimal
consumption rate as the function of the transaction cost $\lambda$
\[
\rho=\frac{\delta}{1-\gamma}-\frac{\gamma}{2}x_0^2+
\frac{\gamma}2 \biggl(\frac{3}4{\frac{x^2_0(1-x_0)^2}{1-\gamma
}}
\biggr)^{2/3}\lambda^{2/3} +O(\lambda).
\]
What probably is surprising here is that the dependence on the actual
state of the process only enters into the $O(\lambda)$ term and the
impatience rate does not show up in the first correction term. Also the
correction in the \emph{relative} consumption rate is positive or
negative depending on the sign of $\gamma$.

\begin{appendix}\label{app}
\section*{Appendix}

\subsection{Free boundary value problem}
In this section, we deal with the resolvability of (\ref{eqode}), where
$\tmu/\tsigma$ and $\rho$ satisfy (\ref{eqlnrho40}) and (\ref{eqrho}).

Both $\rho$ and $\tmu/\tsigma$ are linear expressions of $\tsigma$ with
function coefficients
\begin{eqnarray*}
\frac{\tmu}\tsigma&=&\alpha\tsigma-\alpha\gamma,
\\
\rho&=& \frac{\delta}{1-\gamma}+ \frac{\gamma\alpha}{\gamma-1} \biggl(\mu-\frac{1}2+
\frac{1}2 \bigl((1-\alpha )\tsigma+\alpha\gamma \bigr) \biggr)
\\
&=&-\frac{\alpha(1-\alpha)\gamma}{2(1-\gamma)}\tsigma+\frac
{1}{1-\gamma} \biggl(\delta-\gamma\alpha
\biggl(\mu-\frac{1}2 \biggr)-\frac{\gamma
^2}{2}\alpha^2
\biggr).
\end{eqnarray*}

Dividing by $\tsigma^2$, equation (\ref{eqode}) takes the form
\begin{eqnarray*}
-\frac{1}2 \alpha(1-\alpha)^2 \biggl(\frac{1}{\tsigma}
\biggr)' &=&\frac{1-\alpha}{\tsigma^2} \biggl( \biggl(\frac\tmu\tsigma-
\frac{\tsigma}2 \biggr)- \biggl(\mu-\frac{1}2 \biggr) \biggr) - \biggl(
\frac{1}{\tsigma^2}-\frac{1}{\tsigma^3} \biggr)\rho.
\end{eqnarray*}
So for the function
%
\begin{equation}
\label{eqf-def} f(x)=\frac{1}{\tsigma(x)}
\end{equation}
we have the ODE on $\real\setminus\{0,1\}$
\begin{equation}
\label{eqf} \tfrac{1}2 f'=a_0 f + (1-f)
\bigl((a_1+a_2) f + a_3 f^2
\bigr),
\end{equation}
where
%
\begin{eqnarray}
\label{eqa0a1} (1-\alpha) \biggl( \biggl(\alpha-\frac{1}2 \biggr)
\tsigma- \gamma\alpha- \biggl(\mu- \frac{1}2 \biggr) \biggr) &=&-\alpha(1-
\alpha)^2 \bigl(a_0\tsigma+( \tsigma-1)a_1
\bigr),\hspace*{-25pt}
\\
\label{eqa2a3} \rho&=&\alpha(1-\alpha)^2(a_2
\tsigma+a_3).
\end{eqnarray}
That is, the coefficients can be written,
as
%
\begin{eqnarray}
\label{eqadef} %
 a_0&=&\frac{1}{\alpha(1-\alpha)} \biggl(\mu-
\frac{1}2+\gamma\alpha- \biggl(\alpha-\frac{1}2 \biggr) \biggr)=
\frac{\mu}{\alpha(1-\alpha)}-\frac{1-\gamma}{1-\alpha},\nonumber
\\
a_1&=&-\frac{1}{\alpha(1-\alpha)} \biggl(\mu-\frac{1}2+\gamma\alpha
\biggr),
\nonumber\\[-8pt]\\[-8pt]\nonumber
a_2&=&-\frac{\gamma}{2(1-\gamma)}\frac{1}{1-\alpha},
\\
a_3&=&\frac{1}{(1-\alpha)^2(1-\gamma)} \biggl(\frac{\delta}{\alpha}- \gamma \biggl(
\mu-\frac{1}2 \biggr)-\frac{\gamma
^2}2\alpha \biggr).\nonumber
\end{eqnarray}
All the coefficients $a_0,a_1,a_2,a_3$ are locally Lipschitz continuous
on $\real\setminus\{0,1\}$, and, therefore, the right-hand side
of the ODE (\ref{eqf}) is locally Lipschitz continuous on
$(\real\setminus\{0,1\})\times\real$.
Standard results in ODE theory implies that (\ref{eqf}) is locally uniquely
solvable on $\real\setminus\{0,1\}$, that is, for each
$(x,y)$ there is a
neighborhood $\cU$ of $x$ and a function $f\dvtx \cU\to\real$ such that $f(x)=y$
and $f$ satisfies (\ref{eqf}). Any local solution extends
uniquely
to a maximal connected solution.
Also the solutions do not cross each other, that is, if $f_1$, $f_2$
are two
solutions both defined on an interval $\cU$ and $f_1(x)<f_2(x)$
for some $x\in\cU$
then $f_1<f_2$ everywhere on $\cU$. It also gives that if $f$ is a
solution of some connected set $I\subset
\real\setminus\{0,1\}$ and $f(x)\neq0$ for
some $x\in I$ then $f\neq0$ on $I$.

To construct a shadow price, we need a special solution $f$ to (\ref{eqf}),
defined on some set $I$.
\begin{longlist}[(iii)]
\item  $f$ solves equation (\ref{eqf}) on $I\setminus
\{0,1\}$
and when $0,1\in
I$ then $f(x)$ can be extended continuously to $I$.
\item The boundary condition $g'\mid_{\partial I}=0$
corresponds to
$\tsigma\mid_{\partial I}=1$, that is, \mbox{$f\mid_{\partial I}=1$}.
\item The other requirement for constructing a
shadow price is that the range of $g$ is
$[\ln(1-\ulambda),\ln(1+\olambda)]$. In terms of $\tsigma$, and
$f$ this
requires that
%
\begin{equation}
\label{eqdeltag} \ln \biggl(\frac{1+\olambda}{1-\ulambda} \biggr)= \biggl\llvert \int
_I g'(x) \,dx \biggr\rrvert = \biggl\llvert
\int_I\frac{f(y)-1}{y(1-y)} \,dy \biggr\rrvert,
\end{equation}
since $g'(x)=(1-1/\tsigma(x))/(x(1-x))$.
\item Finally, we also need that $\rho\geq0$.
So $\rho\mid_{\partial I}\geq0$ has to hold. As on $\tsigma\mid
_{\partial
I}=1$,
we obtain a necessary condition, namely
%
\begin{equation}
\label{eqa2+a3>0} \alpha(a_2+a_3)\mid_{\partial I}> 0.
\end{equation}
\end{longlist}

%
\begin{definition}\label{defFBP}
We call the pair $(I,f)$ the solution of the free boundary problem if it
fulfills \textup{(i)--(iv)}.
\end{definition}

Besides the conditions listed above a solution of the free boundary
value is
useful for constructing a shadow price if $g$ obtained from it is strictly
monotone. As $g'$ will be defined from $\tsigma$ by the formula (\ref
{eqgfromtsigma}), a sufficient condition of the monotonicity of $g$ is that
%
\begin{equation}
\label{eqgmon} \qquad\frac{1}{\alpha(1-\alpha)} (f-1)>0\quad\mbox{or}\quad \frac{1}{\alpha(1-\alpha)}
(f-1)<0\qquad\mbox{in the interior of $I$}.
\end{equation}

We reformulate Proposition
\ref{proprho>0} in terms of $f$.

\begin{proposition}\label{propa2+a3f>0} Let $I$ be an interval
and $f\dvtx I\to\real$ a nowhere vanising continous function. Assume that
$f$ solves
(\ref{eqf}) on $I\setminus\{0,1\}$,
\[
\int_I \biggl\llvert \frac{f(z)-1}{z(1-z)} \biggr\rrvert
\,dz< \infty,
\]
and $\alpha(a_2+a_3f) (x)>0$ for some $x\in I$.
Then $\alpha(a_2+a_3f)>0$ on $I$.
\end{proposition}

\begin{pf}
We can define $\tsigma(x)=1/f(x)$ then $\tmu,\rho$ by
(\ref{eqtmurhokappa}). Then $\tsigma,\tmu,\rho$ satisfies
(\ref{eqODEagain}) as
(\ref{eqf}) is only a recasting of this equation. Note that the
coefficients of~(\ref{eqf}) was defined in such a way that
\[
\rho=\alpha(1-\alpha)^2 \biggl(\frac{a_2}f+a_3
\biggr)\qquad\mbox{on } I.
\]

Proposition~\ref{proprho>0} applies to $\tsigma,\tmu,\rho$ and
yields that $\rho$ does not change sign on $I$. The same is true for
$f$ as we already noted. Hence, the sign of $\alpha(a_2+a_3f)$ is also
constant on $I$ and this is the claim.
\end{pf}

The function $a_0$ plays the crucial role in the analysis;
it is
%
\begin{equation}
\label{eqa1+a2+a3} a_0(x)=\frac{\mu}{x(1-x)}-\frac{1-\gamma}{1-x}.
\end{equation}
It turns out to be crucial as $a_0(x_0)=0$ gives a degenerate
solution of the free boundary problem, namely $I=\{x_0\}$,
$f(x_0)=1$. It corresponds to the frictionless case
$\ulambda=\olambda=0$ and $x_0$, usually called the Merton proportion,
is the proportion of the wealth the investor tries to keep in shares.

In what follows, we search for the solution of the free
boundary value problem, such that $x_0$ is in the interior of $I$.
Working out the expression $(1-\gamma)\alpha(1-\alpha)^2(a_2+a_3)$,
we get
\[
\delta-\alpha\gamma \biggl(\mu-\frac{1}2(1-\gamma)\alpha \biggr)=
\delta+ \frac\gamma{2(1-\gamma)} \bigl( \bigl((1-\gamma)\alpha-\mu
\bigr)^2- \mu^2 \bigr).
\]
Observe that the minimum of this function is attained at $x_0$.

So a sufficient condition for (iv) to hold is that
%
\begin{equation}
\label{eqdeltabig} \delta\geq\frac{1}2\frac{\gamma}{1-\gamma}\mu^2.
\end{equation}
Note that if we are interested in the solution of the free boundary
value problem for all sufficiently small transaction costs then
(\ref{eqdeltabig}) is also necessary, provided that the interval
$I$ on which the solution is defined is shrinking onto $x_0$ as the
transaction costs goes to zero. So our standing assumption in the rest
of this section is that~(\ref{eqdeltabig}) is fulfilled.

Let
\begin{eqnarray*}
H_+&=& \bigl\{x\in\real\setminus\{0,1\}\dvtx a_0(x)>0 \bigr\},
\\
H_-&=& \bigl\{x\in\real\setminus\{0,1\}\dvtx a_0(x)<0 \bigr\},
\end{eqnarray*}
and call
$x_0=x_0(\mu,\gamma)=\mu/(1-\gamma)$.
There are the following cases:
\begin{longlist}[(2)]
\item[(1)] $0<(1-\gamma)<\mu$ then
$H_+=(0,1)\cup(x_0,\infty)$, with $x_0>1$,
\item[(2)] $0<\mu=(1-\gamma)$ then $H_+=(0,\infty)$,
\item[(3)] $0<\mu<(1-\gamma)$ then
$H_+=(0,x_0)\cup(1,\infty)$ with $x_0\in(0,1)$,
\item[(4)] $\mu=0$ then $H_+=(1,\infty)$,
\item[(5)] $\mu<0$ then $H_+=(x_0,0)\cup(1,\infty)$, with
$x_0<0$.
\end{longlist}

Not all cases are equally interesting, for example, $\mu=0$ means that
the price~$S$ is a martingale, while $\mu<0$ corresponds to the strict
super-martingale case. In some cases, the optimal strategy on a
frictionless market using the price $S$ does not involve trading,
apart from the initial re-balance of the portfolio. So in these cases
the transaction cost are irrelevant. These are (2) and
(4), that is, when the Merton proportion $\mu/(\gamma-1)$ is
0 or 1. In these two cases, the free boundary value problem has no solution.
Nevertheless, the remaining three cases can be handled in
a similar manner.

In the rest of this section, we will use the following notation.
For $x\in\real\setminus\{0,1\}$, denote by $f_x$ the maximal
connected
solution of (\ref{eqf}) which satisfies $f(x)=1$ and the domain of
$f_x$ by
$\cD_x$. Then $\cD_x$ is a connected open subset of
$\real\setminus\{0,1\}$. We set
\begin{eqnarray*}
s(x)&=& \cases{ \displaystyle\sup\{t\in\cD_x\dvtx{f_x}
\mid_{(x,t)}>1\},&\quad $x\in H_+$,
\vspace*{3pt}\cr
\displaystyle\sup\{t\in
\cD_x\dvtx{f_x}\mid_{(x,t)}<1\},&\quad $x\in
H_-$}
\end{eqnarray*}
and
\[
\cI(x)= \biggl\llvert \int_x^{s(x)}
\frac{f_x(z)-1}{z(1-z)}\,dz \biggr\rrvert.
\]

First, we prove an easy asymptotic result.

\begin{theorem}\label{thma}
Let $\delta>0$ and $x_0=\frac{\mu}{1-\gamma}$ as above.
If $x_0\notin\{0,1\}$ and (\ref{eqdeltabig}) holds
then the free boundary problem has a solution provided that
$\ln\frac{1+\olambda}{1-\ulambda}$ is positive and sufficiently small.
\end{theorem}

In the proofs below, we usually write equation (\ref{eqf}) as
\[
f'(y)=F \bigl(y,f(y) \bigr).
\]
We will use the fact that when $I$ is an interval not contiguous to
$\{0,1\}$ and $J\subset(0,\infty)$ is a bounded interval, then
$F$ is Lipschitz continuous in its second variable on $I\times J$. As
a result, the solution
starting from within $I\times J$ can be continued until it exits from
$I\times J$. When $J=(0,M)$ then we can also note that $F(y,m)/m$ is
bounded on $I\times J$ and therefore any solution must be strictly
positive on such an $I$.

Another fact used frequently below is the following. Take a sequence
$x_n\to x$ such that $x_n\in I$ where $I$ is not contiguous to
$\{0,1\}$ and a bounded interval~$J$. Assume that
$f_{x_n}$ defined on $I$ and
${f_{x_n}}\mid_I$ takes values in $J$ then $f(y)=\lim_{n\to\infty}
f_{x_n}(y)$, $y\in I$ solves (\ref{eqf}) and equal to ${f_x}\mid_{I}$.

\begin{pf*}{Proof of Theorem \ref{thma}}
The function $f_{x_0}$ has a local extremum at $x_0$, since
$f'_{x_0}(x_0)=0$ and $f''_{x_0}(x_0)=a_0'(x_0)\neq0$ by direct
computation. Then take an
interval $\cU$ such that $f_{x_0}\mid_\cU$ has an extremum at $x_0$
and for all $y\in\cU$ the function
$f_y$ is defined on $\cU$. To see this, take a rectangle
$\cU\times J$ which contains $(x_0,1)$ in its interior and such that
$\cU$ is not contiguous to $\{0,1\}$ and $J$ is bounded. Then
$F$ is bounded on $\cU\times J$. Then by decreasing $\cU$,
we can achieve that $\sup_{\cU\times J} \llvert F\rrvert  \leq\llvert J\rrvert /\llvert \cU\rrvert $
where~$\llvert\cdot \rrvert $ denotes the length of the interval. For $x\in
\cU$,
$f_x$ is defined on $\cU$ and $f_x\mid_\cU$ takes values in $J$.

Then there is a left neighborhood $(y_0,x_0)$ of $x_0$ contained in
$\cU$ such
that for $y\in(y_0,x_0)$ we have
$s(y)\in\cU$.
By the continuous dependence of $f_y$ on the parameter~$y$,
we have that $\cI$ restricted to $(y_0,x_0)$ is continuous and obviously
$\cI(y)\to0$ as $y\to x_0$ from the left. So the range
$\{\cI(y)\dvtx y\in(y_0,x_0)\}$
contains $(0,\varepsilon)$, a right neighborhood of 0, for some
$\varepsilon>0$.

For a given $\ln\frac{1+\olambda}{1-\ulambda}<\varepsilon$, we
can find
$y\in(y_0,x_0)$ such that $\cI(y)=\ln\frac{1+\olambda}{1-\ulambda
}$ and take
$([y,s(y)],f_y)$ as the solution of the free boundary value problem.
\end{pf*}

%
\begin{theorem}\label{thmb}
Suppose that $\delta>0$, (\ref{eqdeltabig}) holds and
%
\begin{eqnarray}
\label{eqmu<1-gamma} 0&<&\mu<(1-\gamma),
\\
\label{eqa3>0} \inf_{x\in(0,1)}(1-x)^2
a_3(x)&>&0.
\end{eqnarray}

Then for any $\ulambda\in(0,1)$ and $\olambda>0$, the free boundary problem
has a solution $(I,f)$, $I\subset(0,1)$
is a compact interval, $f>1$ in the interior of $I$.
\end{theorem}

The condition (\ref{eqa3>0}) may be written in terms of the
parameters $\delta,\gamma,\mu$ as follows. Since $x(1-x)^2a_3(x)$ is a
second-order polynomial in $x$ and the leading coefficient is negative
it is positive in on $(0,1)$ exactly when it is positive at $0$ and at
$1$. Its value at $0$ is $\delta/(1-\gamma)>0$ so the condition is
that it is positive at $1$ which gives
%
\begin{equation}
\label{eqdeltabig2} \delta>\gamma \bigl(\mu-\tfrac{1}2(1-\gamma) \bigr).
\end{equation}

\begin{pf*}{Proof of Theorem \ref{thmb}}
The proof is similar to that of Proposition~4.2 in~\cite{MR2676941}.

Here $H_+=(0,x_0)\cup(1,\infty)$ with $x_0\in(0,1)$. Below we use the
notation introduced
before Theorem~\ref{thma}. We show that:
\begin{longlist}[(a)]
\item[(a)] $(x,1)\subset\cD_x$ for $x\in(0,x_0)$,
\item[(b)] $s,\cI$ are continuous on $(0,x_0)$ and
$s(x)<1$ for $x\in(0,x_0)$.
\item[(c)] $f_x>1$ on $(x,s(x))$,
\item[(d)] $\lim_{x\to0+}\cI(x)=\infty$, $\lim_{x\to
x_0-}\cI(x)=0$.
\end{longlist}
Then there is an $x\in(0,x_0)$ such that
$\cI(x)=\ln\frac{1+\olambda}{1-\ulambda}$ and we take $I=[x,s(x)]$.
Then the pair $(I,f_x)$ solves the fr
free boundary problem in the above sense,
\mbox{(i)--(iii)} is obvious and (iv) follows
from (\ref{eqdeltabig}) as we have seen.

For (a), we borrow an idea from Kallsen and Muhle-Karbe \cite{MR2676941}. We actually
show that $f_x$ cannot break out from bounded interval on $(x,1)$. This
guarantees that the solution can be continued on the whole half-line
$(x,1)$.

To see boundedness, note that
$(1-\alpha)a_0,(1-\alpha)a_1,(1-\alpha)a_2$ are bounded on $[x,1)$ and
$\inf_{(0,1)} (1-\alpha)^2a_3>0$ by assumption so $a_3$ determines
the main term in
$F(y,M)$ for $M$ large.
More precisely,
there is a
threshold $M_0>1$, such that
%
\begin{equation}
\label{eqbounded} F(y,M) < 0\qquad\mbox{for }y\in(x,1)\mbox{ and } M\geq
M_0.
\end{equation}
Then\vspace*{1pt} $\sup_{y\in[x,1)} f_x(y)< M_0$. Indeed, $y_1<1$ with
$y_1=\inf\{y\geq\in[x,1)\dvtx f_x(y)\geq M_0\}$ would immediately
yield a
contradiction as $f'_x(y_1)$ should be both negative and nonnegative. This
proves~\textup{(a)}.

For $s(x)<1$, note that
$(1-\alpha)^2a_1,(1-\alpha)^2a_2$ both tend to zero
at 1. On the other hand, $\inf_{(0,1)} (1-\alpha)^2 a_3>0$ by
assumption and
$\lim_{y\to1^-} (1-y) a_0(y)<0$ as $\mu<(1-\gamma)$. This
implies that
there is $\eta>0$ and a threshold $y_0>0$ such that
%
\begin{equation}
\label{eqeta} F(y,M) < -\frac{\eta}{1-y}\qquad\mbox{for }y
\in[y_0,1)\mbox{ and } M\geq1.
\end{equation}
By (\ref{eqeta}), $f'_x(y) < -\eta/(1-y)$ for $y_0<y<s(x)$ and
\[
1-f(y_0) \leq f(y)-f(y_0)\leq\eta\ln \biggl(
\frac{1-y}{1-y_0} \biggr)\qquad\mbox {for }y_0<y<s(x)
\]
gives that $s(x)$ cannot be one.
Whence $s(x)<1$ and $\cI(x)$ is finite
since $f_x$ is continuous on $[x,s(x)]$.

As we have seen for $y\in(0,1)$, the mapping $x\mapsto f_x(y)$ is
continuous on $(0,1)$. From this, the continuity of $x\mapsto s(x)$
follows as $f'_x(s(x))\neq0$ for $x\in(0,1)$.
Using the dominated
convergence theorem, we also obtain the continuity of $\cI$.

\textup{(c)} is obvious: $x\in H_+$ so $f_x'(x)>0$,
$f_x(x)=1$, so on $(x,s(x))$ the function $f$ is positive by the
definition of $s(x)$.

The second half of (d), that is, $\lim_{x\to
x_0}\cI(x)=0$ is just the continuity of
$\cI$.
To show that $\lim_{x\to0} \cI(x)=\infty$, we use that near zero the
main term on the right-hand side of (\ref{eqf}) is $a_0f$.
Taking $\eta> 0$ small enough, this leads to the
existence a threshold $y_0\in(0,x_0)$ such that
%
\begin{equation}
\label{eqsmallx} F(y,M) >\frac{\eta}{y}\qquad\mbox{for } y
\in(0,y_0)\mbox{ and }1\leq M <1+\eta.
\end{equation}
We get $\lim_{x\to0^+}\cI(x)=\infty$ from (\ref{eqsmallx}) by the following
reasoning. For a given $y>0$,
the limit $\overline{f}(y)=\lim_{x\to0^+} f_x(y)$ exists, as
$x\mapsto f_x(y)$ is decreasing in $x$.
Now $\overline{f}(y)\leq1+\eta$ would lead to $\eta\geq
f_x(y)-1=\int_x^y
f'_x(z)\,dz>\int_x^y \eta/z\,dz$ for all $0<x<y$, a contradiction.
Hence, $\lim_{x\to0^+}f_x(y)\geq1+\eta$ for all
$y\in(0,y_0)$.
Then
\[
\liminf_{x\to0^{+}} \cI(x)\geq \lim_{x\to0^+}\int
_x^{y_0} \frac{f_x(y)-1}{y(1-y)}\,dy\geq\eta \int
_0^{y_0}\frac{1}{y(1-y)}\,dy=\infty.
\]
So it is enough to prove (\ref{eqsmallx}). Note that
$\lim_{x\to0}xa_0(x)=\mu>0$ and\break $\lim_{x\to0}xa_i(x)$, $i=1,2,3$ are
bounded. So for small $y_0,\eta$ the effect of
\[
y \bigl\llvert (1-M) \bigl( \bigl(a_1(y)+a_2(y)
\bigr)M+a_3(y)M^2 \bigr) \bigr\rrvert \leq 3 \eta(1+
\eta)^2 \max \bigl(\llvert  a_1\rrvert,\llvert a_2\rrvert,\llvert a_3\rrvert \bigr) (y)
\]
is negligible compared to $ya_0(y)M>\eta$ for any $M\in(1,1+\eta)$
and $y\in(0,y_0)$. This yields (\ref{eqsmallx}).
\end{pf*}

Similar analysis applies to the other two nondegenerate cases.

\begin{theorem}\label{thmc}
Suppose that $\delta>0$, both (\ref{eqdeltabig}), (\ref
{eqdeltabig2}) hold and
%
\begin{equation}
\label{eq1-gamma<mu} 1-\gamma<\mu.
\end{equation}

Then for any $\ulambda\in(0,1)$ and $\olambda>0$, the free boundary problem
has a solution $(I,f)$, with $I\subset(0,\infty)$ and
$(1-f)/(1-\alpha)$ is continuous on $I$ and negative in the interior
of $I$.
\end{theorem}

Here, the difficult case is when $1$ belongs to the interior of
$I$. The continuity of $(1-f)/(1-\alpha)$ at $1$ shows that even in
this case $f'$ is continuous on $I$.
\begin{pf*}{Proof of Theorem \ref{thmc}}
In this case, $H_+=(0,1)\cup(x_0,\infty)$ with
$x_0=\mu/(1-\gamma)>1$. As before, we denote by $f_x$ the maximal
connected solution of~(\ref{eqf}) such that $f_x(x)=1$.

We consider $f_x$ for $x\in(0,1)\cup(1,x_0)$.
We show below that
the next properties
hold for $f_x$, $s$ and $\cI$:
\begin{longlist}[(a)]
\item[(a)] $[x,s(x))\subset\cD_x$, and $f_x<1$ on
$(x,s(x))$ for $x\in(1,x_0)$,
\item[(b)] $s(x)<\infty$ and $f_x(s(x))=1$ for $x\in(1,x_0)$,
\item[(c)] $s(x)=1$ and $\lim_{y\to1^-} f_x(y)=1$
for $x\in(0,1)$,
\item[(d)] $\cI$ is finite valued and continuous on
$(0,1)\cup(1,x_0)$,
\item[(e)] $\lim_{x\to0^+}\cI(x)=\infty$,
$\lim_{x\to x_0-}\cI(x)=0$, $\lim_{x\to1^-}\cI(x)=0$.
\end{longlist}
Taking these properties for granted,
if $\lim_{x\to1^+} \cI(x)>\ln(\frac{1+\olambda}{1-\ulambda})$
then there is an $x\in(1,x_0)$ such that
$\cI(x)=\ln(\frac{1+\olambda}{1-\ulambda})$
and with $I=[x,s(x)]$ the pair $(I,f_x)$ is a solution of the free boundary
value problem as in the proof of Theorem~\ref{thmb}. Now, $f_x<1$ in the
interior of $I$ by~\textup{(a)}.

However,\vspace*{1pt} it is also possible that
$\cI(\infty)\leq\ln(\frac{1+\olambda}{1-\ulambda})$.
Then the solution
is constructed from two components; the first one is the limit of the above
solutions as the initial point $x$ tends to one from right.

So, we let $f_{1}(y)=\lim_{x\to1^+\hspace*{-0.5pt}}
f_x(y)$ for $y\in(1,s(1))$, where
$s(1)=\lim_{x\to1^+} s(x)$. The function $f_{1}$ and the
point $s(1)$ is well
defined as $x\mapsto f_x(y)$ is a increasing function of $x$ for each fixed
$y\in(1,s(1))$ while $x\mapsto s(x)$ is decreasing on
$(1,x_0]$.

Then $f_{1}$ solves (the integral version) of
(\ref{eqf}), hence continuous and $f_{1}$ is also a solution of
(\ref{eqf}). We also denote by $f_{1}$ the maximal connected
solution extending~$f_{1}$. Then:
\begin{longlist}[(a)]
\item[(f)] $s(1)<\infty$, $f_{1}(s(1))=1$,
\item[(g)] $\lim_{y\to1^+} f_{1}(y)=1$,
\item[(h)]
\[
\int_{1}^{s(1)} \biggl\llvert
\frac{f_{1}(z)-1}{z(1-z)} \biggr\rrvert \,dz
=\lim_{x\to1^+} \cI(x). %
\]
%
\message{^^Jea=\meaning\ea^^J}
\end{longlist}

So, when $\cI(1)<\ln(\frac{1+\olambda}{1-\ulambda})$ there is an
$x_2\in(0,1)$ such that
%
\begin{equation}
\label{eqintf}
\int_{x_2}^{1} \biggl\llvert
\frac{f_{x_2}(z)-1}{z(1-z)} \biggr\rrvert \,dz
+
\int_{1}^{s(1)} \biggl\llvert
\frac{f_{1}(z)-1}{z(1-z)} \biggr\rrvert \,dz =\ln \biggl(\frac{1+\olambda}{1-\ulambda} \biggr).
\end{equation}
Then we take $I=[x_2,s(1)]$ and $f=f_{x_2}\cup
f_{1}$ and conclude that the pair
$(I,f)$ is the solution of the free boundary value problem in the
sense of Definition~\ref{defFBP}. Indeed, \textup{(i)--(ii)} are
clear, for (iii) we have to add that $(f(z)-1)/(z(1-z))$
does not change sign on $I$ so (\ref{eqintf}) implies (\ref{eqdeltag}),
while for (iv) follows from (\ref{eqdeltabig}).

It remains to prove the properties listed above.

For $1<x<x_0$, as $x\in H_{-}$, we get immediately that $f_x<1$ on
$(x,s(x))\cap\cD_x$. So $f_x$ is
defined on $[x,s(x))$. We conclude that \textup{(a)} holds.

For \textup{(c)}, note that for $x\in(0,1)$ the solution $f_x$ is
bounded. The proof is identical to the one given in Theorem~\ref
{thmb}, as
it only used that $\inf_{(0,1)} (1-\alpha)^2a_3>0$ which holds by assumption
(\ref{eqdeltabig2}). So $f_x$, $x\in (0,1)$ is defined on $[x,1)$ and since
$(0,1)\subset H_{+}$ $f_x(y)=1$ for $y\in(x,1)$ is impossible. This yields
$s(x)=1$ for $x\in(0,1)$, which is the first part of \textup{(c)}.

We prove below that
%
\begin{eqnarray}
\label{eqs-inf} s(1)&<&\infty,
\\
\label{eqfxlim} \lim_{y\to1^-} \frac{f_x(y)-1}{1-y}&=&\lim
_{y\to1} \frac
{(1-y)a_0(y)}{(1-y)^2a_3(y)}\qquad\mbox{for }x\in(0,1),
\\
\label{eqf-inflim} \lim_{y\to1^+} \frac{f_{1}(y)-1}{1-y}&=&\lim
_{y\to1}\frac{(1-y)a_0(y)}{(1-y)^2a_3(y)}.
\end{eqnarray}

Equation~(\ref{eqs-inf}) implies \textup{(f)} by the continuity of the
function $f_{1}$. Since the solutions do not cross each other,
$x\mapsto
s(x)$ is decreasing on $(1,x_0)$, which combined with
(\ref{eqs-inf}) yields~\textup{(b)}.

Equation~(\ref{eqfxlim}) and (\ref{eqf-inflim}) give that $\cI$ is finite
valued on $(0,x_0)$.
As the solution $f_x$ depends continuously on $x$, we obtain the continuity
of $\cI$ also both on $(0,1)$ and on $[1,x_0)$. So (\ref{eqfxlim}) and
(\ref{eqf-inflim}) imply~\textup{(d)},
\textup{(g)}, \textup{(h)} and the second part of~\textup{(c)}.
Moreover, they also imply the
finiteness of the integral of $(f_x(z)-1)/(z(1-z))$ for $x\in(0,1]$,
which yields
the last two
relations of~\textup{(e)}. For the first limit, $\lim_{x\to0^+}
\cI(x)=\infty$, the end of the proof of Theorem~\ref{thmb} applies
as it
only used that $\lim_{x\to0^+} ya_0(y)=\mu>0$, which holds by
(\ref{eq1-gamma<mu}).

So the proof is completed by showing (\ref{eqs-inf}), (\ref{eqfxlim})
and (\ref{eqf-inflim}).
\end{pf*}

\begin{pf*}{Proof of (\ref{eqf-inflim})}
We use that $\lim_{y\to1} (1-y)a_i(y)$ exists for $i=0,1,2$,
and $\lim_{y\to1}(1-y)^2 a_3(y)$ exists and positive by the
assumption (\ref{eqdeltabig2}).
This implies that for $y$ near $1$ the dominating term of
$(1-y)^2F(y,M)$ is $(1-M)M^2 a_3$, for any $M>0$.
That is,
\[
\lim_{y\to1} \frac{(1-y)^2F(y,M)}{1-M}>0\qquad\mbox{for }M>0.
\]
Then
for each $M\in(0,1)$ there
exists $\eta=\eta(M),\varepsilon=\varepsilon(M)>0$ such that
\[
F(y,M)> \varepsilon\qquad\mbox{for }1<y\leq1+\eta.
\]
This proves that for $1<x\leq y\leq1+\eta$, the relation
$f_x(y)>M$, since otherwise
$y_0=\inf\{y>x\dvtx f_x(y)=M\}\leq1+\eta$ and $f_x(y_0)=M$,
$f'_x(y_0)=F(y_0,M)>0$ would lead to a contradiction. But then
for $1<y<1+\eta(M)$ we have $f_1(y)=\lim_{x\to1^+}f_x(y)\geq M$. Since
this is true for all $M\in(0,1)$, we obtained that $\lim_{y\to1} f_{1}(y)=1$.

The limit $\lim_{y\to1}
(1-y)a_0(y)=\mu-(1-\gamma)>0$ by (\ref{eq1-gamma<mu}).
Then for the function $h(y)=f_1(1+1/y)$ we have that
\begin{eqnarray*}
h'(y)&=&-\frac{1}{y^2}f'_1(1+1/y)
\\
&=& -\frac{2}{y^2} a_0f_1(1+1/y)+ \bigl(1-h(y) \bigr)
\frac{-2}{y^2} \bigl((a_1+a_2)f_1+a_3f_1^2
\bigr) (1+1/y)
\\
&=& \frac{a(y)}{y}+ \bigl(1-h(y) \bigr)b(y),
\end{eqnarray*}
where $a(y)=-\frac{2}{y}(a_0f_1)(1+1/y)$ and
$b(y)=\frac{-2}{y^2}((a_1+a_2)f_1+a_3f^2_1)(1+1/y)$. Both $a,b$ has a
limit as $y\to\infty$, we have $a(\infty)=\lim_{y\to\infty}
a(y)=2(\mu-(1-\gamma))>0$ and $b(\infty)=\lim_{y\to\infty} b(y)<0$.
There is $y_0$ and $\eta>0$ such that $b(y)<-\eta$ for $y>y_0$ and
rearrangement gives that
\[
\bigl( \bigl(1-h(y) \bigr)e^{-\eta y} \bigr)'= \biggl(-
\frac{a(y)}{y}+ \bigl(1-h(y) \bigr) \bigl(\eta-b(y) \bigr)
\biggr)e^{-\eta
y}.
\]
Integrating both sides between $y$ and $\infty$ and multiplying by
$ye^{-\eta y}$, we get
\[
y \bigl(1-h(y) \bigr) = y\int_y^\infty \biggl(
\frac{a(z)}{z}- \bigl(1-h(z) \bigr) \bigl(\eta -b(z) \bigr)
\biggr)e^{-\eta(z-y)}\,dz.
\]
First we estimate from above the nonnegative quantity $y(1-h(y))$
\[
\limsup_{x\to\infty} y \bigl(1-h(y) \bigr) \leq\limsup
_{y\to\infty}y\int_y^\infty \biggl(
\frac{a(z)} {z} \biggr)e^{-\eta(z-y)}\,dz\leq\frac{a(\infty
)}\eta<\infty.
\]
Using this estimation we get that
\begin{eqnarray*}
\biggl\llvert y \bigl(1-h(y) \bigr)-\frac{a(\infty)}{\eta} \biggr\rrvert &\leq&
\int_0^\infty \biggl\llvert \frac{a(y+z)y}{y+z}-a(
\infty) \biggr\rrvert e^{-\eta
z}\,dz
\\
&&{} + \int_0^\infty y\bigl(1-h(y+z) \bigr) \bigl(
\eta-b(y+z) \bigr)e^{-\eta z}\,dz.
\end{eqnarray*}
Here, the first term is small provided that $y$ is sufficiently
large, while the second term is small if $\eta$ is close to
$b(\infty)$ and $y$ is large. In summary, we obtained that
\[
\lim_{y\to1^+}\frac{1-f_1(y)}{y-1}= \lim_{y\to\infty}
y \bigl(1-h(y) \bigr)=\frac{a(\infty)}{b(\infty)}.
\]
\upqed
\end{pf*}

\begin{pf*}{Proof of (\ref{eqfxlim})}
The proof is very similar to that of
(\ref{eqf-inflim}). First, the limits $\lim_{y\to1}(1-y)^2a_i(y)$,
$i=0,1,2,3$ exist, equal zero for $i=0,1,2$ and positive for
$i=3$.
So for any $\eta>0$
there is $\varepsilon>0$ and a threshold $y_0\in(0,1) $ such that
\[
F(y,M) < -\frac{\eta}{(1-y)^2}\qquad\mbox{for }y\in(y_0,1)\mbox{ and
}M>1+\eta.
\]
As $[x,1)\subset\cD_x$ this yields that $\limsup_{y\to1^-}
f_x(y)\leq1+\eta$. This is true for all $\eta>0$, so we have
$\lim_{y\to1^-} f_x(y)=1$.

We refine this estimation similarly as above by taking
$h(y)=f_x(1-1/y)$. Then
\[
h'(y)=\frac{1}{y^2} f'_x(1-1/y)=
\frac{a(y)}{y}+ \bigl(1-h(y) \bigr)b(y),
\]
where $a(y)=\frac{2}y (a_0f_x)(1-1/y)$ and
$b(y)=\frac{2}{y^2}((a_1+a_2)f_x+a_3f^2_x)(1-1/y)$. Both $a,b$ has a limit
at $\infty$, $a(\infty)=\lim_{z\to1^-} (1-z)a_0(z)>0$ and
$b(\infty)=\lim_{z\to1} (1-z)^2 a_3(z)>0$. Then we get for
$0<\eta<b(\infty)$ that
\[
\bigl( \bigl(h(z)-1 \bigr)e^{\eta z} \bigr)'=
\frac{a(z)}z e^{\eta z} + \bigl(h(z)-1 \bigr) \bigl(\eta-b(z) \bigr).
\]
Since $h(1/(1-x))=f_x(x)=1$, we get by integrating, now from $1/(1-x)$
to $y$ and multiplying with $ye^{-\eta y}$ that
\begin{eqnarray*}
y \bigl(h(y)-1 \bigr)&=& y\int_{1/(1-x)}^y \biggl(
\frac{a(z)}{z}+ \bigl(h(z)-1 \bigr) \bigl(\eta-b(z) \bigr)
\biggr)e^{-\eta(y- z)} \,dz.
\end{eqnarray*}
As $\lim_{z\to\infty}(h(z)-1)(\eta-b(z))<0$, we get by substituting
$z=y-r$ that
\[
\limsup_{y\to\infty} y \bigl(h(y)-1 \bigr)\leq\lim
_{y\to\infty} \int_0^{y-1/(1-x)} a(y-r)
\frac{y}{y-r} e^{-\eta r} \,dr=\frac{a(\infty
)}{\eta}.
\]
Then we compare $y(h(y)-1)$ to
$a(\infty)/\eta$ as above and obtain (\ref{eqfxlim}).
\end{pf*}

\begin{pf*}{Proof of (\ref{eqs-inf})}
Since $f_{1}(y)\to1$ as $y\to1$ and $\inf_{y\in(1,(1+x_0)/2)}
(a_2+a_3)(y)>0$,
we have that there is $x$ such that $a_2(x)+a_3(x)f_{1}(x)>0$. Let $f$
be a maximal connected solution of (\ref{eqf}) such that
$0<f(x)<f_{1}(x)$, but $a_2(x)+a_3(x)f(x)>0$ still holds. Denote by
$s=\sup\{y\in\cD(f)\dvtx f(y)<1\}$.

On $(x,s)$, we have $0<f<1$, hence the solution can be
continued on this whole interval, that is, $(x,s)\subset\cD(f)$, where
$\cD(f)$ is the domain of $f$.

Since the solution does not cross each other,
we have $s(1)<s$ and it is enough to show that $s<\infty$.

Assume on the contrary that
$s=\infty$. Using Proposition~\ref{propa2+a3f>0} and the fact
that $\alpha(x)=x>0$ on $(1,\infty)$, we obtain that $a_2+a_3f>0$
on $\cD(f)$. Then
\[
f' \geq a_0 f+(1-f)f a_1=(a_0+a_1)
f-f^2 a_1.
\]
Note that $\lim_{x\to\infty} (1-x)a_1(x)=-\gamma$. So the sign of
$a_1$ near $\infty$
depends on the sign of $\gamma$.
If $a_1> 0$ in a neighborhood of $\infty$ then even $f'\geq a_0 f$
holds, in the opposite case we use the estimate $f'\geq(a_0+a_1)f$.
Note that
\begin{eqnarray*}
\lim_{y\to\infty} (1-y) (a_0+a_1) (y)&=&-1,
\\
\lim_{y\to\infty} (1-y)a_0(y)&=&-(1-\gamma).
\end{eqnarray*}
So there is a threshold $y_0>x_0 $ and $\eta>0$ such that
\[
f'(y)\geq\frac{\eta}{\llvert y-1\rrvert } f(y)\qquad\mbox{for
}y>y_0.
\]
But then $f$ cannot be bounded and we obtained a contradiction.
\end{pf*}


\begin{theorem}\label{thmd}
Suppose that $\delta>0$, $\mu<0$ and (\ref{eqdeltabig}).

Then for any $\ulambda\in(0,1)$ and $\olambda>0$, the free boundary problem
has a solution $(I,f)$, $I\subset(-\infty,0)$ is a compact or
semi-closed interval;
in the later case, the open end point is $0$. Finally, $f<1$ in the
interior of $I$.
\end{theorem}

\begin{pf}Since $\mu<0$, $x_0=\mu/(1-\gamma)<0$ and
$H_+=(x_0,0)\cup(1,\infty)$.
For $x\in(-\infty,x_0)$, we take $f_x$ the maximal
connected solution with $f_x(x)=1$.
Then we show that:
\begin{longlist}[(a)]
\item[(a)] $(x,s(x))\subset\cD_x$ and $f_x(s(x))=1$
when $s(x)<0$,
\item[(b)] $\cI$ is finite valued and continuous on
$(-\infty,x_0)$,
\item[(c)] $\lim_{x\to-\infty} \cI(x)=\infty$ and
$\lim_{x\to x_0-} \cI(x)=0$.
\end{longlist}
Then one can find $x\in(-\infty,x_0)$ such that
$\cI(x)=\ln(\frac{1+\olambda}{1-\ulambda})$.
If $s(x)<0$ then we take
$J=[x,s(x)]$, otherwise $J=[x,0)$. The
pair $(I,f_x)$ solves the free boundary value problem in the sense of
Definition~\ref{defFBP}. Cases~\textup{(i)--(iii)} obviously hold, while
for (iv) follows from (\ref{eqdeltabig}).

Only $\lim_{x\to-\infty} \cI(x)=\infty$ requires justification, all
other properties are clear from the definitions. We proceed as at the
end of
the proof of Theorem~\ref{thmb}. In this case, $\lim_{y\to-\infty}
ya_i(y)$ exists and equal zero for $i=2,3$, while the limit is
finite for $i=0,1$, especially $\lim_{y\to-\infty} ya_0(y)=(1-\gamma)>0$.
It easily follows that there is a threshold
$y_0<x_0$ and a positive $\eta$ such that
%
\begin{equation}
\label{eqF<etay} F(y,M) < -\frac{\eta}{\llvert y\rrvert }\qquad\mbox{for }y<y_0
\mbox { and }1-\eta\leq M\leq1.
\end{equation}
This implies that $\lim_{x\to-\infty} f_x(y)\leq1-\eta$ and $\cI
(x)\to\infty$
and $x\to-\infty$.
\end{pf}
\end{appendix}

\section*{Acknowledgments}
We thank the referees for their comments which improved the
presentation of the result. After submitting the first
version of this paper, we learned about the work of
Choi, S{\^{\i}}rbu and {\v{Z}}itkovi{\'c} \cite{2012arXiv12040305C}. They independently and concurrently developed
similar ideas, pushing it even further than we did,
with more emphasis on the optimal control approach.



%

\printaddresses
\end{document}